\documentclass[12pt]{JHEP3}
\usepackage{graphicx,amsmath,amssymb}
\usepackage{subfigure}
\allowdisplaybreaks

\usepackage{bm,amsmath,amssymb}

\def\bg{\begin{eqnarray}}
\def\nd{\end{eqnarray}}

\title{Thermodynamics of large N gauge theory from top down holography}

\author{Mohammed Mia \\
Columbia University, New York, 10027, USA\\
\vskip.07in
{\tt  mm3994@columbia.edu }}

\abstract{By considering fluxes on $D7$ branes and explicitly computing their back-reaction on the geometry with and without a black hole, 
we show how the UV divergence of Klebanov-Strassler model can be regulated. Using the form of the metric and fluxes in the extremal and
non-extremal limit,  we compute the on-shell gravity action including the localized sources up to linear order in perturbation parameter
$\frac{g_sM^2}{N}, g_sN_f$ where $N,M$ and $N_f$ are units of $D3,D5$ and $D7$ charges in the dual gauge theory. Using the gravitational description, we show how the gauge
theory undergoes a first-order
Hawking-Page like phase transition and  compute the critical temperature $T_c$. 
Finally, we obtain the equation of state for the 
gauge theory by computing thermodynamic state functions of the black hole and exhibit how black holes in deformed cone
geometry can lead to results that are
qualitatively similar to lattice QCD simulations.}

\maketitle

\begin{document}
\section{Introduction}
 Asymptotic freedom guarantees that at high temperatures, nuclear matter is best described as a weakly interacting gas of
quarks and gluons. The weak nature of the coupling allows a perturbative description of QCD where observables in principal
 can be
computed with increasing accuracy as temperature is increased. However at low temperatures, nuclear matter is 
color neutral, indicating color degrees of
freedom are strongly coupled and  confined inside hadrons. The strong coupling regime of QCD is analyzed by either studying
the theory on the lattice or resorting to effective field theories- both of which have their success and limitations. 

On the other hand, gauge theories naturally arise from excitations of open strings ending on branes \cite{Witten:1995im}
 while gravitons can be
described by excitations of closed strings. By studying the
interaction between open and closed strings, one can relate the Hilbert space of the gauge theory with that of gravity. The
best studied example is the AdS/CFT correspondence \cite{Mal-1}: Here  the gauge theory is four dimensional 
maximally SUSY ${\cal N}=4$
 conformal field theory and
the gravitons describe a ten dimensional space $AdS_5\times S^5$ i.e. five dimensional anti-deSitter space times a compact
five sphere. When the t'Hooft coupling for the gauge theory is large, Hilbert space of the gauge theory is conjectured to be
contained in the Hilbert space of gravitons described by classical action of  $AdS_5\times S^5$ geometry.
Thus a strongly coupled quantum gauge theory gets a classical description in terms of weakly coupled 
gravity. Expectation values in gauge theory which are otherwise extremely difficult to compute due to strong coupling, can 
easily be computed using the dual holographic description \cite{Witt-1}. 
   
Since QCD is a gauge theory, the obvious question becomes is there a holographic description for a QCD like theory?
 Any model that attempts to mimic QCD should feature  
it's two key attributes: deconfinement at high temperatures
and confinement at low temperatures. Thus a holographic description of QCD, if it exists, should incorporate both the
deconfined and confined phase while allowing us to study the thermodynamics near the critical region where
 phase transition occurs. Since QCD coupling is large near the critical temperature $T_c$, the corresponding 'tHooft
 coupling is also large and  we expect the holographic 
 description
  to be most accurate near $T_c$. While for $T\gg T_c$,  QCD coupling is small rendering  pQCD techniques to be most
  reliable and we do not require a holographic description. 
 In addition, lattice QCD simulations suggest that the conformal anomaly is largest near $T_c$ 
 \footnote{Analysis of high temperature phase and phase 
transitions  can be found in
\cite{Gross:1980br}-\cite{Pisarski:1983ms}. For recent developments in lattice QCD, please consult 
\cite{Petreczky:2012rq}-\cite{Borsanyi:2012ve} and \cite{Panero:2009tv}-\cite{Panero:2012qx} for the large N limit.}. 
 Thus we should look for a holographic description near $T_c$ where the gauge coupling is large and the 
 theory is highly non-conformal.

  The AdS/CFT correspondence only considers conformal field theories, where there is no phase
 transition and no critical temperature $T_c$. Black holes in $AdS_5\times S^5$, describing  a thermal CFT
  can  mimic large $T\gg T_c$
 regime of QCD. But as already mentioned, at large $T$ we can simply use perturbative QCD to study the thermodynamics quite
 accurately. The regime near $T\sim T_c$ where pQCD breaks down is where
 holographic techniques can be most valuable. But since black holes in AdS space cannot describe a non-conformal theory, 
 we must find a generalization of AdS/CFT correspondence to describe thermal gauge theories that undergo phase transitions.

 Obtaining a geometric description of a gauge theory that resembles thermal QCD near phase transition is a formidable task.
 Before attempting to find a holographic map between QCD and gravity, we must first explore the general scope of gauge/gravity
 correspondence and understand how gauge theories with non-trivial Renormalization
 Group (RG) flows arise in string theory. In principal   
 excitations of  D branes placed  in various  geometries give rise to gauge theories. The fluxes and scalar
 fields sourced by the branes back-react and warp the geometry. This warped geometry is referred as the `dual geometry'. The
 fluxes and dilaton field in the dual geometry are used to obtain the RG flow of the gauge theory. While all the
 thermodynamic state functions of the gauge theory can be obtained by identifying the
 partition function of the dual geometry with that of the gauge theory.

 Since QCD is non supersymmetric and non-conformal, the first objective is to find gauge theories with RG flows
  arising from D brane configurations with minimal
 SUSY. A great deal of progress has been made in that direction: In
 \cite{9904017, 9906194} RG flows that connected conformal fixed points at IR and UV was incorporated, 
and \cite{9909047}  connected the 
UV ${\cal N} = 4$ conformal fixed point to a ${\cal N} = 1$ confining theory. But the model with QCD like 
RG flow 
 and 
minimal supersymmetry   
is the Klebanov-Strassler (KS)
model \cite{KS} which was obtained by considering IR modifications to Klebanov-Tseytlin (KT) model \cite{Klebanov:2000nc} (with an extension \cite{Ouyang} to incorporate fundamental matter by considering D7 branes). However,
 at the highest energies  the gauge theory is nothing like QCD:  the effective degrees of freedom diverges and the
  gauge theory is best described in terms of bifundamental fields.  
Only at the lowest energies the gauge theory resembles ${\cal N}=1$ SUSY QCD which confines. 
 In the limit when effective $D3$ brane charge is
large, the gauge theory with large 'tHooft coupling has an equivalent description in terms of warped deformed cone. We
can learn about this gauge theory which is very different from QCD in the UV, using the dual description .   

 In a series of papers \cite{FEP}-\cite{Mia:2012ue}, we proposed the general procedure to modify the UV dynamics of KS
 theory. In this paper we demonstrate how UV
 modifications are realized with an exact calculation of fluxes and metric as follows:  
We first consider world volume fluxes $\widetilde{F}_2$ on D7 branes embedded in KS geometry with or without a black hole.
These fluxes $\widetilde{F}_2$ induce anti-D3 and anti-D5 charges such that the total effective $D5$ charge vanishes, 
while the effective $D3$ charge no longer diverges in the far UV. This way, the UV divergence of KS theory is removed. 
The resulting dual geometry takes the form of 
 a warped deformed cone at small radial distances and $AdS_5\times T^{1,1}$ far away from the tip of the cone. Thus, we have
 confinement at IR, dual to deformed cone at small radial distance and conformal gauge theory at UV, dual to AdS space.

It is worth mentioning the great deal of effort given in computing the black hole geometry in the presence of flux and scalar fields. 
For example in \cite{KT-non-ex, PandoZayas:2006sa, thorimal, Caceres:2011zn} the cascading picture of the original KS model
was extended to incorporate black-hole without any fundamental matter, while fundamental matter was accounted for in 
\cite{cotrone}. However, since these black holes are obtained in KS geometry, the dual gauge theories are UV
divergent and quite distinct from QCD.  

Additionally, most of the attempts are based on obtaining an effective lower dimensional action from 
KK reducing ten dimensional 
supergravity action. Dimensional reduction of a generic ten dimensional action  can be quite  challenging
 specially when there are
non-trivial fluxes and scalar fields. It is also highly non-trivial to obtain a consistent truncation.
 Furthermore, 
it is not clear how RG flow of the dual gauge theory can be obtained since the fields in the effective action are not 
the dilaton or the flux in the original ten dimensional action. 

In our approach, we directly work with the ten dimensional geometry,
avoiding the difficulty of KK reduction. With the UV divergences of KS theory removed, we study the thermodynamics of the 
gauge theory by directly identifying the gauge theory partition function 
with that of the ten dimensional geometry.   
The thermal gauge theory arising from the brane excitations
 has a rich phase structure and as temperature is altered, we expect phase transitions.
 In this work, we make progress in that direction and    
   obtain ten dimensional geometry (with or without a black hole) that arises from low energy limit
of type IIB superstring theory including localized sources. 

The role of localized sources is crucial in our analysis, since
they allow us to modify the geometry at large radial distances. They also give rise to a radial scale $r_0$ and 
consequently
an energy scale $\Lambda_0$: Warped geometry in small $r$ region i.e. $r<r_0$ correspond to IR modes $(\Lambda<\Lambda_0)$
 of the gauge theory while
inclusion of large $r$ region i.e. $r>r_0$ correspond to including UV modes $(\Lambda>\Lambda_0)$ of the gauge theory.   
 For any given temperature of the dual gauge theory, 
there are two geometries$-$ extremal (without black
hole) and non-extremal (with black hole) but the geometry with lower on-shell action is preferred. 
At a critical temperature $T_c$,
 both geometries are equally likely and we have a phase transition. We evaluate critical temperature
  using a perturbative analysis and $T_c$
 depends on the boundary conditions as well as the scale $r_0$. Thus the localized sources directly influence the
 thermodynamics of the gauge theory.
 
 An alternative approach to construct gravitational description of gauge theories is to start with non-critical string theory and
 consider the resultant dual geometry \cite{Klebanov:2004ya}. For a QCD like gauge theory that confines in the IR and becomes
 free in the UV, one can obtain a five dimensional dual geometry \cite{kirit}. The
  gravity action includes dilaton field and an
 effective potential for the dilaton. However, the geometry has large curvature and
higher order terms in $\alpha'$ need to be included, which will modify the classical gravity action. By considering part of the
higher order terms, one can find an effective dilaton potential which in turn can reproduce the QCD beta function.
In a bottom up scenario, this effective potential is tuned to fit lattice QCD results for the conformal anomaly and the Polyakov
loop. Since the potential is not derived directly from an underlying brane configuration, there is no guarantee that the 
geometry is in fact a holographic image of a gauge theory. 

On the other hand, in our top down approach we proceed by 
first analyzing
brane excitations  in conifold geometries where the field theory has  global and local symmetries common to that of QCD. 
At low energies, the excitations give rise to a four dimensional gauge theory which decouples from gravity and can be described 
holographically by the 
low energy limit of critical superstring theory i.e supergravity in ten dimensions. As we study a gauge theory arising from strings 
ending on branes, we know the field content of the theory and in some cases, the exact superpotential at zero temperature. 
Hence our top down approach is
distinct from the bottom up models where a precise knowledge of the gauge theory is lacking or the gravity action is incomplete.
 To make meaningful quantitative 
 comparisons with QCD, one  must identify the gauge theory for which the dual gravity is being constructed. 
 While in bottom up models this
 identification is not clear,  in our top down approach, it
 is automatic. 
 Thus the
 phenomenology that results from this gravity description can be directly compared to that of QCD as the gauge theory 
 resembles large N QCD.

 The paper is organized as follows: In section 2.1, we obtain exact values for type IIB fluxes in warped ten dimensional 
 geometry in the presence of $D7$ branes both
 in extremal (no black hole) and non-extremal (black hole) limit. The metric and fluxes are evaluated as a Taylor series 
  in perturbative
 parameter ${\cal O}\left(\frac{g_s M^2}{N}\right), {\cal O}\left(g_sN_f\right)$
  where $N,M$ and $N_f$ are number of $D3,D5$ 
 and $D7$ branes in the dual gauge theory. For the metric, terms up to linear
 order are evaluated while the fluxes are obtained at zeroth order. In section 2.2 we propose a brane configuration that
  can source such fluxes and demonstrate how UV divergence of KS theory can be removed. Using the metric and flux, in section 3.1 the on-shell gravity action is exactly
 evaluated up to linear order and Hawking-Page like transition is analyzed. In section 3.2, the effect of localized
 source is incorporated and thermodynamic state functions are obtained. Finally in section 3.3 connections to QCD are
 established by considering small black holes in deformed cone geometry.

\section{Gauge/String duality: From branes to geometry}
As already mentioned in the introduction, holographic map between gauge theory and gravity can be constructed by studying
excitations of branes placed in certain geometries. The gauge theory arises from open strings ending on the branes while the
interactions between open and closed strings leave a holographic imprint of the gauge theory on the geometry. This imprint is
captured by the warped dual geometry. At the lowest energies, open and closed string sector decouples and we are left with a
gauge theory living in flat four dimensional space which can be described by the dual geometry.  

The dual geometry has a classical action with fluxes and localized sources. The classical action is enough to describe the
geometry, since the curvature will be small everywhere which in turn can be guaranteed by considering large $M$. In the following section, we
analyze this classical action and then in section 2.2, we will describe the brane configuration that can give rise to such
geometry.  
\subsection{Geometry: Fluxes and localized sources in type IIB theory}
Consider the type IIB action including $N_f$ number of coincident Dp branes in string frame\footnote{For $N_f\neq 1$, the
abelian action we wrote down gets modified and we need to consider the non-abelian action \cite{Tseytlin:1997csa}.
Approximating the non-abelian action by taking $N_f$ copies of the abelian action means that we are not distinguishing between 
different flavors and ignoring their interactions. For our purpose, we can simply set $N_f=1$ and consider
the abelian case.}:
\bg \label{Actionscom}
S_{\rm total}&=&S_{\rm SUGRA}+N_f\;S_{Dp}\nonumber\\
S_{\rm SUGRA}&=&\frac{1}{2\kappa^2_{10}}\int d^{10}x \sqrt{G^s}\Bigg(e^{-2\phi}\left(R_s+4(\triangledown \phi)^2\right)-
\frac{F_1^2}{2}
-\frac{|\widetilde{F}_5|^2}{4\cdot 5!}-\frac{G_3 \cdot \bar{G}_3}{12}\Bigg)\nonumber\\
&+&
\frac{1}{8i\kappa_{10}^2}\int \frac{C_4\wedge G_3\wedge \bar{G}_3}{{\rm Im} \tau}
\nd 
where $\tau=C_0+ie^{-\phi}$, $F_1=dC_0$ and  $G^s={\rm det}g_{MN}^s, M,N=0,..,9$, $g_{MN}^s$ is the metric in string frame and 
$G_3=F_3-\tau H_3$.
 Here the action for a $p$ brane, upto quadratic order in flux $\widetilde{F}_{ab}\equiv B_{ab}+F_{ab} $ is given
 by\footnote{We will be considering $D7$ branes for which the Chern-Simons action contains $\sim \mu_7\alpha'^2 \int
 \left[C_4\wedge tr(R\wedge R)-e^{-\phi} tr(R\wedge \ast R)\right]$ where $R$ is the pullback of the curvature two form. 
 However we consider geometries such that $R\wedge R$ and $R\wedge \ast R$ is zero on the world volume of the brane. 
 } 
\bg \label{Action1s}
S_{Dp}=-\int d^{p+1}\sigma T_p\sqrt{-f^s}\left(1+\frac{1}{4} \widetilde{F}^{ab}\widetilde{F}_{ab}\right)+ \mu_p\int \left(C e^{\widetilde{F}}\right)_{p+1}
\nd
Here $f^s={\rm det} f_{ab}^s$, $f_{ab}^s=g_{MN}^s \partial_a X^M \partial_b X^N$ is the pull back metric,
$B_{ab}=B_{MN} \partial_a X^M \partial_b X^N$, $B_{MN}$ is NS-NS two form and $C_{p+1}$ is the RR flux. Also,
$\widetilde{F}_{ab}$ is raised or lowered with the pullback metric $f_{ab}^s$ in string frame.

The action (\ref{Actionscom}) is complex due to the topological term $\sim -iC_4\wedge G_3\wedge {\bar G}_3$. By taking the real part
of the action and minimizing it, we can obtain the real valued fluxes $F_5,F_3,H_3$, the metric $g_{MN}^s$ and the scalar fields $\phi,C_0$.
We will consider the following real action
\bg\label{Actions}
S_{\rm SUGRA}&=&\frac{1}{2\kappa^2_{10}}\int d^{10}x \sqrt{G^s}\Bigg(e^{-2\phi}\left(R_s+4(\triangledown \phi)^2\right)-
\frac{F_1^2}{2}
-\frac{|\widetilde{F}_5|^2}{4\cdot 5!}-\frac{G_3 \cdot \bar{G}_3}{12}\Bigg)\nonumber\\
&-&
\frac{1}{4\kappa_{10}^2}\int C_4\wedge H_3\wedge F_3
\nd 
   
We can simplify equations resulting from variation of the above action by absorbing the scalar field in the definiton of the
metric. This is done by going to the Einstein frame defined through $g_{MN}=g_{MN}^se^{-\phi/2}$, where the action (\ref{Actions})
takes the following form 
\bg \label{Action}
S_{\rm SUGRA}&=&\frac{1}{2\kappa^2_{10}}\int d^{10}x \sqrt{G}\left(R+\frac{\partial_M
\tau\partial^M\bar{\tau}}{2|{\rm
Im}\tau|^2}-\frac{|\widetilde{F}_5|^2}{4\cdot 5!}-\frac{G_3 \cdot \bar{G}_3}{12 {\rm Im}\tau}\right)\nonumber\\
&-&
\frac{1}{4\kappa_{10}^2}\int C_4\wedge H_3\wedge F_3
\nd 

\bg \label{Action1}
S_{Dp}=-\int d^{p+1}\sigma \;T_p\;e^{\frac{\phi(p+1)}{4}}\;\sqrt{-f}\left(1+e^{-\phi}\frac{1}{4} \widetilde{F}^{ab}\widetilde{F}_{ab}\right)+ 
\mu_p\int \left(C e^{\widetilde{F}}\right)_{p+1}
\nd
where $f={\rm det} f_{ab}$, $f_{ab}=g_{MN} \partial_a X^M \partial_b X^N$ and $\widetilde{F}_{ab}$ is raised or lowered with the pullback
metric $f_{ab}$ in Einstein frame. 
The background warped metric takes
the following familiar form 
\bg\label{metric}
ds^2&=&g_{MN}\;dx^M dx^N\equiv g_{\mu\nu} \;dx^\mu dx^\nu+g_{mn}\; dx^m dx^n\nonumber\\
&=&-e^{2A+2B}dt^2+e^{2A}(dx^2+dy^2+dz^2)+e^{-2A-2B}\tilde{g}_{mn} dx^m dx^n
\nd
where the internal unwarped metric  is given by $\tilde{g}_{mn}\equiv \tilde{g}_{mn}^0+\tilde{g}_{mn}^1 $ with
\bg\label{inmate}
\tilde{g}_{mn}^0 dx^m dx^n&=& \frac{1}{2}{\cal A}^{4/3} K(\rho)\Big[\frac{1}{3K^3(\rho)}\left(d\rho^2+e^{2B}(g^5)^2\right)+{\rm
cosh}^2\left(\frac{\rho}{2}\right)e^{2B}\left[(g^3)^2+(g^4)^2\right]\nonumber\\
&&+ {\rm
sinh}^2\left(\frac{\rho}{2}\right)e^{2B}\left[(g^1)^2+(g^2)^2\right]\Big]\nonumber\\
K(\rho)&=&\frac{\left({\rm
sinh}(2\rho)-2\rho\right)^{1/3}}{2^{1/3}{\rm sinh}\rho}
\nd
Here $\tilde{g}_{mn}^0$ is the metric of the base of deformed cone while $\tilde{g}_{mn}^1 $ is the perturbation due
to the presence of fluxes and localized sources.
Also ${\cal A}$ is a constant, $g^i,i=1,..,5$ are one forms given by 
\bg\label{oneforms}
&&g^1=\frac{e^1-e^3}{\sqrt{2}},~~~~g^2=\frac{e^2-e^4}{\sqrt{2}}\nonumber\\
&&g^3=\frac{e^1+e^3}{\sqrt{2}},~~~~g^4=\frac{e^2+e^4}{\sqrt{2}},~~~g^5=e^5\nonumber\\
&& e^1\equiv-{\rm sin}\theta_1 \;d\phi_1, ~~~~ e^2\equiv d\theta_1\nonumber\\
&& e^3\equiv {\rm cos}\psi \;{\rm sin}\theta_2 \;d\phi_2-{\rm sin}\psi \;d\theta_2,\nonumber\\
&& e^4\equiv {\rm sin}\psi \;{\rm sin}\theta_2\; d\phi_2+{\rm cos}\psi\; d\theta_2,\nonumber\\
&& e^5\equiv d\psi +{\rm cos}\theta_1\; d\phi_1+{\rm cos}\theta_2
\;d\phi_2 \nd 
and   $m.n=4,..,9$ denote the internal `cone' direction while $\mu,\nu=0,..,3$ run over Minkowski directions. 
The warp factor $A(x^m),B(x^m)$ are  functions of
the cone coordinate $x^m=\rho,\psi,\phi_1,\phi_2,\theta_1,\theta_2$. 
Observe that with a
change of coordinates 
\bg \label{rrho} r^3={\cal A}^2 e^\rho \nd
 for large
$\rho$, the metric becomes 
\bg\label{inmate1} &&\widetilde{g}_{mn}^0 dx^m
dx^n\sim dr^2+r^2e^{2B}\left(\frac{1}{9} (g^5)^2+\frac{1}{6}\sum_{i=1}^{4}
(g^i)^2\right) 
\nd 
which is the metric of regular cone with base
$T^{1,1}$. Thus only for small radial coordinate $\rho$, the
internal metric is a deformed cone while at large $\rho$, we really
have a regular cone with topology of $R\times T^{1,1}$.

We will now consider $p=7$, that is we embed D7 branes in the large $\rho$ region where $r$ is the more convenient radial
coordinate. Adding these sources in the large $\rho$ region means that we are only modifying the UV of the dual gauge theory and we
expect that the IR of gauge theory remain mostly unaltered. The effect of this brane embedding for the gauge theory will be
discussed in detail in section 2.2. 

The D7 branes fill up 
Minkowski space
$(t,x,y,z)$, stretching along $r$ direction and filling up $S^3$ inside the $T^{1,1}=S^3\times S^2$. We consider two branches:

\vskip.1in
\noindent $\bullet$  Branch I  with parametrization
$(\sigma^0,\sigma^1,..,\sigma^7)=(t,x,y,z,r,\psi,\phi_2,\theta_2)$. The brane fills up 4D Minkowski space and stretches along the $r$
direction, filling up an $S^3$ inside $T^{1,1}=S^3\times S^2$. It is a point $(\phi_1(\sigma^\alpha),\theta_1(\sigma^\alpha))$ inside $S^2$
and we pick a profile such that $\theta_1(\sigma^\alpha)=\pi/2$ and $\phi_1(\sigma^\alpha)\equiv \tilde{\phi}_1(r)$ is only a function of the $r$. The
DBI part of the
world volume action for this branch takes the form 

\bg\label{SI}
S_I&=&- |\mu_7|\;\left(V_4\;\int d\Omega^I\; dr\frac{r^3e^{\phi}}{18}\sqrt{1+\frac{e^{2B}r^2}{6}\tilde{\phi}_1'^2}+\int\frac{1}{2} \widetilde{F}_2^I\wedge \ast_f
\widetilde{F}_2^I\right)\nonumber\\
\nd
where $V_4\equiv \int d^4x, d\Omega^I\equiv  d\psi\; d\phi_2\; d\theta_2\; {\rm sin}(\theta_2)$,
 we have denoted world volume flux on branch I with $\widetilde{F}^I_2$ and 
$\ast_f$ is the Hodge star with respect to the pullback metric $f$ of the branch. In obtaining the above action from (\ref{Action1}), we have
used the definition of tension of Dp brane, $T_p=|\mu_p| e^{-\phi}$. Now observe that Branch I of D7 brane is a point on an $S^2$ with volume form 
\bg
\Omega_1\equiv {\rm sin}(\theta_1) d\phi_1\wedge d\theta_1
\nd
Thus the Chern-Simons action for Branch I can be written as 
\bg \label{ScsI}
S_{\rm CS}^I&=& \mu_7\int \Gamma_1\left(C_8+ C_6\wedge
\widetilde{F}_2^I+\frac{1}{4}C_4\wedge \widetilde{F}_2^I\wedge \widetilde{F}_2^I+\frac{1}{6}C_2\wedge \widetilde{F}_2^I\wedge
\widetilde{F}_2^I\wedge \widetilde{F}_2 
\right)\wedge \Omega_1\nonumber\\
\nd
where $C_i$ is the pullback of the RR form to the world volume 
\footnote{In deriving (\ref{ScsI}), we put an additional factor of $1/2$ in front of $C_4$, starting with the Chern-Simons action
(\ref{Action1}).
This is because five form $F_5=dC_4$ is self dual, that is $\ast_{10}dC_4=dC_4$. Then $C_4$ includes both electric and magnetic flux 
since typically
$\ast E=B$, where $E$ and $B$ are electric and magnetic fluxes. Defining $C_4=C_4^{RR}+C_4^{NS}$, we can get $\ast
dC_4^{NS}=dC_4^{RR}$, $\ast
dC_4^{RR}=dC_4^{NS}$ and $dC_4^{RR}=1/2 dC_4,\; dC_4^{NS}=1/2 \ast dC_4$.
The Chern-Simons term is $1/2 \;C_4^{RR}\wedge \widetilde{F}_2\wedge \widetilde{F}_2$, and since $C_4^{RR}=1/2 C_4$, we get the
expansion (\ref{ScsI})} 
\bg
\Gamma_1=\delta(\theta_1-\pi/2) \delta\left(\phi_1-\tilde{\phi}_1(r)\right)
 \nd
 
\vskip.1in
\noindent $\bullet$  Branch II  with parametrization
$(\sigma^0,\sigma^1,..,\sigma^7)=(t,x,y,z,r,\psi,\phi_1,\theta_1)$. Again the brane fills up 4D Minkowski space and stretches along the $r$
direction, fills up an $S^3$ inside $T^{1,1}=S^3\times S^2$ but now is a point $(\phi_2(\sigma^\alpha),\theta_2(\sigma^\alpha))$ inside $S^2$.
Again we pick a profile such that $\theta_2(\sigma^\alpha)=\pi/2$ and $\phi_2(\sigma^\alpha)\equiv \tilde{\phi}_2(r)$ is only a function of the $r$. 
The
DBI part of the
world volume action for this branch takes the form 

\bg\label{SII}
S_{II}&=&-|\mu_7|\;\left(V_4\;\int d\Omega^{II}\; dr \frac{r^3 e^{\phi}}{18}\sqrt{1+\frac{e^{2B}r^2}{6}\tilde{\phi}_2'^2}+\int \frac{1}{2} \widetilde{F}_2^{II}\wedge \ast_f
\widetilde{F}_2^{II}\right)\nonumber\\
\nd
where $d\Omega^{II}\equiv  d\psi\; d\phi_1\; d\theta_1\; {\rm sin}(\theta_1)$.
\begin{table}
\begin{tabular}{|l|p{0.3in}|p{0.3in}|p{0.3in}|p{0.3in}|p{0.3in}|p{0.3in}|p{0.3in}|p{0.3in}|p{0.3in}|p{0.3in}|} \hline
\em {\rm Branch} &$t$ &$x$ &$y$&$z$&$r$&$\psi$&$\phi_1$&$\phi_2$&$\theta_1$&$\theta_2$\\\hline
I &$\surd$&$\surd$&$\surd$&$\surd$&$\surd$&$\surd$&-&$\surd$&-&$\surd$\\
II &$\surd$&$\surd$&$\surd$&$\surd$&$\surd$&$\surd$&$\surd$&-&$\surd$&-\\\hline
\end{tabular}
\end{table} 
Observing that Branch II of D7 brane is a point on an $S^2$ with volume form 
\bg
\Omega_2\equiv {\rm sin}(\theta_2) d\phi_2\wedge d\theta_2
\nd
the Chern-Simons action for Branch II can be written as 
\bg \label{ScsII}
S_{\rm CS}^{II}&=& \mu_7\int \Gamma_2\left(C_8+ C_6\wedge
\widetilde{F}_2^{II}+\frac{1}{4}C_4\wedge \widetilde{F}_2^{II}\wedge \widetilde{F}_2^{II}+\frac{1}{6}C_2\wedge \widetilde{F}_2^{II}\wedge
\widetilde{F}_2^{II}\wedge \widetilde{F}_2^{II} 
\right)\wedge \Omega_2\nonumber\\
\nd
where  
\bg
\Gamma_2=\delta(\theta_2-\pi/2) \delta\left(\phi_2-\tilde{\phi}_2(r)\right)
 \nd
Note that $C_2$ has no legs in the $t,x,y,z$ direction and thus the last term in both (\ref{ScsI}) and (\ref{ScsII}) do not
contribute.  Now varying the action (\ref{Action1}) with respect to $\widetilde{F}_2$, we get the following equation 
\bg\label{F2f}
d\left(\ast_f \widetilde{F}_2\right)=d\left[\frac{2\mu_7}{|\mu_7|} \left(C_6+\frac{1}{2}C_4\wedge \widetilde{F}_2\right)\right]
\nd
where $\widetilde{F}_2=\widetilde{F}_2^I$ or $\widetilde{F}_2^{II}$.

 Note that $G_3=F_3-\tau H_3$, and $\ast_{10} F_3= dC_6+I_7$ where $I_7$ is some seven form. Then the action 
 $S_{\rm total}$ (where $S_{\rm SUGRA}$ is given by (\ref{Action})) to be
 stationary under variation  
 $C_6$ and $C_2$ gives
 
 \bg \label{G3}
 &&d\left(\frac{\widetilde{F}_3}{{\rm Im}\tau}\right)= 4\kappa_{10}^2\mu_7N_f\left(\Gamma_1\widetilde{F}_2^{I}\wedge \Omega_1+\Gamma_2\widetilde{F}_2^{II}\wedge \Omega_2\right)\nonumber\\
 &&d\left(\frac{\ast_{10} \widetilde{F}_3}{{\rm Im}\tau}- C_4\wedge H_3\right)=0\nonumber\\
 \nd
 where $\widetilde{F}_3=F_3-C_0H_3$. In deriving the above equation from (\ref{Action}) and (\ref{Action1}), we used the relation
 $G_3\cdot \bar{G}_3=|\widetilde{F}_3|^2+|{\rm Im\tau} H_3|^2$ and $\int \sqrt{G}|\widetilde{F}_3|^2=\int 6\widetilde{F_3}\wedge
 \ast_{10}\widetilde{F_3}$.

 Now observe that $\widetilde{F}_5=dC_4+I_5$, where $I_5$ is some five form.  Then for the action to be stationary under
  variation
  of $C_4$ gives the Bianchi identity 
 \bg \label{BI}
 d\widetilde{F}_5=H_3\wedge F_3-\mu_7\kappa_{10}^2 N_f\left(\Gamma_1 \widetilde{F}^I\wedge \widetilde{F}^I\wedge 
 \Omega_1+
 \Gamma_2 \widetilde{F}^{II}\wedge \widetilde{F}^{II}\wedge \Omega_2\right)
 \nd
  Now Lorentz invariance and self-duality fixes the five form to be
 \bg \label{F5}
 \widetilde{F}_5=\left(1+\ast_{10}\right) d\alpha \wedge dt\wedge dx \wedge dy \wedge dz 
 \nd 
 where $\alpha(x^m)$ is a scalar function.  Variation of the action with respect to the space time metric $g_{MN}$  gives the Einstein equations
 
 \bg \label{Ricci_Min}
R_{\mu\nu}&=&-g_{\mu\nu} \left[\frac{G_3 \cdot \bar{G_3}}{48\; {\rm
Im}\tau}+\frac{\widetilde{F}_5^2}{8\cdot 5!}\right]+\frac{\widetilde{F}_{\mu
abcd}\widetilde{F}_\nu^{\;abcd}}{4 \cdot 4!}
+\kappa_{10}^2N_f \left(T_{\mu\nu}^{\rm loc}-\frac{1}{8} g_{\mu\nu} T^{\rm loc}\right)\nonumber\\
R_{mn}&=&-g_{mn} \left[\frac{G_3 \cdot \bar{G_3}}{48 \;{\rm
Im}\tau}+\frac{\widetilde{F}_5^2}{8\cdot 5!}\right]+\frac{\widetilde{F}_{m
abcd}\widetilde{F}_n^{\;abcd}}{4 \cdot 4!}+\frac{G_m^{\;bc}\bar{G}_{nbc}}{4\;{\rm Im}\tau}
+\frac{\partial_m\tau \partial_n \tau}{2\;|{\rm Im}\tau|^2}\nonumber\\
&+&\kappa_{10}^2N_f \left(T_{mn}^{\rm loc}-\frac{1}{8} g_{mn} T^{\rm
loc}\right) 
\nd
where $\tilde{R}_{mn}$ is the Ricci tensor for the metric $\tilde{g}_{mn}$ and $T_{MN}^{\rm loc}$ is defined through
\bg
 T^{\rm loc}_{MN}= -\frac{2}{\sqrt{G}}\frac{\delta S_{D7}}{\delta g^{MN}} 
 \nd   
 We want to solve the flux equations (\ref{F2f}),(\ref{G3}), (\ref{BI}), the Einstein equations (\ref{Ricci_Min}) and simultaneously find the embedding 
 $\tilde{\phi}_i(r)$ that minimizes the
 action. But before we do so, observe that $\Omega_i$ is closed form, i.e. $d\Omega_i=0$. Then taking derivative of the first equation in (\ref{G3}) gives
 \bg
 \Gamma_1d\left({\rm Im}\tau\widetilde{F}_2^{I}\right)\wedge \Omega_1+\Gamma_2 d\left({\rm Im}\tau\widetilde{F}_2^{II}\right)\wedge \Omega_2=0
 \nd
 which has the solution 
 \bg \label{dF2}
 d\left({\rm Im}\tau \widetilde{F}_2^{I}\right)=0, ~~~~~~~~~ d\left({\rm Im}\tau\widetilde{F}_2^{II}\right)=0
 \nd 
On the other hand, the four form is given by 
 \bg\label{C4-ex}
C_4&=&\alpha(x^m) \;dt\wedge dx\wedge dy \wedge dz
\nd
The scalar function $\alpha(x^m)$ can be obtained from (\ref{BI}), which is the Bianchi identity while the warp factor $A,B$
can be obtained from the Einstein equations. We now proceed as follows: 
The Ricci tensor in the Minkowski direction takes
the following simple form
\bg \label{ricci_Min}
R_{\mu\nu}&=&-\frac{1}{2} \left[\partial_m (g^{mn} \partial_n
g_{\mu\nu})+ g^{mn} \Gamma^M_{nM}\partial_m g_{\mu\nu}-
g^{mn}g^{\nu'\mu'}\partial_m g_{\mu'\mu}
\partial_n g_{\nu'\nu}\right]
\nd
where $\nu',\mu'=0,..,3$ and $\Gamma^M_{nM}$ is the Christoffel
symbol. Now using the ansatz (\ref{metric}) for the metric,
(\ref{ricci_Min}) can be written as

\bg\label{ricci_Min_a}
R_{tt}&=&e^{4(A+B)}\left[\widetilde{\triangledown}^2(A+B)-3\widetilde{g}^{mn}\partial_nB\partial_m(A+B)\right]\nonumber\\
R_{ij}&=&-\eta_{ij}
e^{2(2A+B)}\left[\widetilde{\triangledown}^2A-3\widetilde{g}^{mn}\partial_nB\partial_mA\right]
\nd where the Laplacian is defined as
\bg\label{Laplacian}
\widetilde{\triangledown}^2=\widetilde{g}^{mn}\partial_m\partial_n
+\partial_m\widetilde{g}^{mn}\partial_n+\frac{1}{2}\widetilde{g}^{mn}\widetilde{g}^{pq}\partial_n\widetilde{g}_{pq}
\partial_m
\nd

The set of equations can be simplified by taking the trace of the first equation in (\ref{Ricci_Min}) and using
(\ref{ricci_Min_a}). Doing this we get
\begin{eqnarray} \label{warp_eq_1}
\widetilde{\triangledown}^2(4A+B)-3\widetilde{g}^{mn}\partial_nB\partial_m(4A+B)
&=&
e^{-2A-2B}\frac{G_{mnp}\bar{G}^{mnp}}{12\textrm{Im}\tau}+e^{-10A-4B}\partial_m\alpha\partial^m\alpha
\nonumber\\
&&+\frac{k^2_{10}N_f}{2}e^{-2A-2B}(T^m_m-T^{\mu}_{_\mu})^{loc}
\end{eqnarray}
 On the other hand  using
(\ref{Ricci_Min}) in (\ref{ricci_Min_a}), one gets
\bg \label{BHfactorA1}
R_t^t-R_x^x=0
\nd
which in turn would immediately imply
\bg\label{BHfactorA}
\widetilde{\triangledown}^2 B-3\widetilde{g}^{mn} \partial_m B \partial_n B=0
\nd
Now let's look at the Bianchi identity. Using (\ref{F5}) in
(\ref{BI}) gives
\begin{eqnarray}\label{bi}
\tilde{\triangledown}^2\alpha -3e^{-2A-2B}\partial_m
B\partial^m\alpha&=&e^{2A-B}\frac{\ast_6 \bar{G}_3\cdot
G_3}{12i\textrm{Im}
\tau}+2e^{-6A-3B}\partial_m e^{4A+B}\partial^m\alpha+ {\cal L}_{\rm loc}\nonumber\\
\end{eqnarray}
where $\ast_6$ is the Hodge star for the metric $g_{mn}$ and 
\bg\label{local}
{\cal L}_{\rm loc}\sim \mu_7\kappa_{10}^2N_f \widetilde{F}^{ab}\widetilde{F}_{ab} e^{4A}\Gamma_i
\nd

Now taking the trace of the first equation in (\ref{Ricci_Min}), subtracting it from
(\ref{bi}), we get
\begin{eqnarray}\label{GKP_BH}
&&\widetilde{\triangledown}^2\left[e^{4A}\left(e^{B}-\gamma\right)\right]=\frac{e^{2A-B}}{24\textrm{Im}\tau}|\textrm{i}
G_3-\ast_6G_3|^2+e^{-6A-3B}|\partial e^{4A}\left(e^{B}-\gamma\right)|^2\nonumber\\
&&+3e^{-2A-2B}\partial_mB\partial^m\left[e^{4A}\left(e^{B}-\gamma\right)\right]+\frac{k^2_{10}N_f}{2}e^{2A-B}(T^m_m-T^{\mu}_{_\mu})^{loc}
-{\cal L}_{\rm loc}
\end{eqnarray}
where $\gamma\equiv \alpha e^{-4A}$. Now, we can
ignore the localized term 
\bg  
 {\cal I}_{\rm loc}= \frac{\kappa^2_{10}N_f}{2}e^{2A-B}(T^m_m-T^{\mu}_{_\mu})^{loc}-{\cal L}_{\rm loc} 
 \nd 
 which is in fact second or higher order
in our perturbation, as we shall see in  what follows. Solving (\ref{warp_eq_1}), (\ref{BHfactorA}) and (\ref{GKP_BH}) together will give
the scalar functions $\alpha, A$ and $B$. We can solve the system perturbatively, order by order in our perturbative parameter 
\bg \label{ppara}
\epsilon \equiv {\cal
O}\left(\frac{g_s\widetilde{M}^2}{N}\right), {\cal O}\left(g_sN_f\right)
\nd
where $\widetilde{M}$ is a unit less constant defined through 
$G_3 \sim \widetilde{M} {\cal G}_3$, ${\cal G}_3$ being a 3-form and $N_f$ is the number of  D7 branes. 

Now writing $\ast_{10} dC_0=2|{\rm Im}\tau|^2 dC_8$, we get the following equation from $S_{\rm total}$, by varying with respect to
$C_8$
\bg
dF_1\sim 4\kappa_{10}^2\mu_7N_f\left(\Gamma_1\Omega_1+\Gamma_1\Omega_2\right)
\nd 
which leads to 
\bg
C_0\sim {\cal O}(N_f) 
\nd
Using this normalization of the axion field, we get following scaling of the dilaton field, using F-theory
\bg\label{Imtau}
\frac{1}{{\rm Im}\tau}=e^{\phi}=g_s\left(1+{\cal O}(g_sN_f) {\cal J}(x^m)\right)
\nd
 where ${\cal J}(x^m)$ is some
function describing the running of dilaton. 
 We solve (\ref{warp_eq_1}), (\ref{BHfactorA}) and (\ref{GKP_BH}) 
 by only 
 considering terms up to ${\cal O}(\epsilon)$ i.e.
 linear order in our perturbation. In the large $\rho$ region, by the switching to $r$ coordinate, we find following 
 scaling of the solution with our perturbative parameter   
\bg \label{alpha}
&&\gamma= 1+ \sum_{l=1}^{\infty}{\cal O}\left(\frac{g_s\widetilde{M}^2}{N}\right) {\cal O}
\left(\frac{\tilde{r}_h}{r}\right)^{4l}\nonumber\\
&&e^{-4A}=\frac{27\pi N\alpha'^2}{4r^4}\left[1+\sum_{j=0}^{\infty}{\cal O}
\left(\frac{\tilde{r}_h}{r}\right)^{4j} {\cal O}\left(\frac{g_s\widetilde{M}^2}{N}\right)\right]\nonumber\\
&&e^{2B}=1-\frac{\tilde{r}_h^4}{r^4}+\sum_{l=1} {\cal O}
\left(\frac{\tilde{r}_h}{r}\right)^{4l} {\cal O}\left(\frac{g_s\widetilde{M}^2}{N}\right)
\nd 
It is instructive to note that at zeroth order in our perturbation, the above solution gives an AdS warp factor and a Schwarzchild black hole
with horizon radius $\tilde{r}_h$. When perturbation is included, the true horizon surface $x^m=x^m_h$ defined through the 
relation $e^{B(x^m=x^m_h)}=0$, would have radial location $r_h$ given by
\bg
r_h=\tilde{r}_h\left(1+{\cal O}\left(\frac{g_s\widetilde{M}^2}{N}\right)\right)
\nd
   
   Using (\ref{alpha}) and the form of $C_4$ as given in (\ref{C4-ex}), equation (\ref{F2f})
can be solved with
\bg \label{F2}
\ast_4 \widetilde{F}_2&=&\frac{\mu_7}{|\mu_7|} \widetilde{F}_2\left(1+ \sum_{l=1}^{\infty}{\cal O}
\left(\frac{g_s\widetilde{M}^2}{N}\right) {\cal O}
\left(\frac{\tilde{r}_h}{r}\right)^{4l}\right)\nonumber\\
C_6&=&\frac{|\mu_7|e^{4A}}{2\mu_7}\left(e^B-1\right) dt\wedge dx \wedge dy \wedge dz\wedge \widetilde{F}_2
\nd
where $\ast_4$ is the Hodge star for the metric $f_{\alpha\beta}, \alpha,\beta\neq 0,1,2,3$. 

Combining (\ref{dF2}) and (\ref{F2}), we see that  $\widetilde{F}_2$ is  self dual (or anti-self dual) while ${\rm Im }\tau \widetilde{F}_2$ is
 {\it closed}  at zeroth order in our perturbation.
 Thus it takes the following form
\bg \label{F2sol}
\widetilde{F}_2^I&=&\left(1+\frac{\mu_7}{|\mu_7|}\ast_4 \right) \frac{{\cal M}\alpha'\sqrt{1+\frac{r^2e^{2B}}{6}\tilde{\phi}_1'^2}}{re^B {\rm Im \tau}}
\left[dr\wedge d\psi +
 a \;dr\wedge d\phi_2\right]\nonumber\\
\widetilde{F}_2^{II}&=&\left(1+\frac{\mu_7}{|\mu_7|}\ast_4\right) \frac{-{\cal M}\alpha'\sqrt{1+\frac{r^2e^{2B}}{6}\tilde{\phi}_2'^2}}{re^B {\rm Im \tau}}
\left[dr\wedge d\psi +
a \;dr\wedge d\phi_1\right]
 \nd 
 where ${\cal M},a$ are constants.
 The factor $(1+\ast_4)$ makes the flux self dual while the function 
 $f_i\equiv \frac{\sqrt{1+\frac{r^2e^{2B}}{6}\tilde{\phi}_i'^2}}{re^{B} {\rm Im \tau} }$ is exactly
 chosen for closure. One can readily check that using $\tilde{g}_{mn}^0$ as the internal metric, indeed the above 
 flux satisfies closure and self duality. 
 
 The form in (\ref{F2sol}) also makes it clear that localized term that we ignored to obtain (\ref{GKP_BH}),  are  
 \bg 
 {\cal L}_{\rm loc}&\sim& {\cal O}\left(\frac{g_s^2N_f{\cal M}^2}{N^2}\right)\nonumber\\
 &\sim&  {\cal O}\left(\frac{g_s{\cal M}^2}{N}\right){\cal O}\left(g_sN_f\right)
 {\cal O}\left( \frac{1}{N}\right)\nonumber\\
 \kappa_{10}^2N_fe^{2A}\left(T^m_m-T^\mu_\mu\right)&=&\kappa_{10}^2N_f|\mu_7|e^{4A}\sqrt{f}\left(|\widetilde{F}_2^I|^2+|\widetilde{F}_2^{II}|^2\right)
 \nonumber\\
 &\sim& {\cal O}\left(\frac{g_s^2N_f{\cal M}^2}{N^2}\right)\sim  {\cal O}\left(\frac{g_s{\cal M}^2}{N}\right){\cal O}
 \left(g_sN_f\right){\cal O}\left( \frac{1}{N}\right)
 \nd 
 Then viewing (\ref{GKP_BH}) as an equation for $\gamma$ gives that the localized terms are of second order in
 our perturbation- which justifies ignoring them.  On the other hand, solving the second equation in (\ref{Ricci_Min}), one obtains  that 
 \bg\label{gmn}
 \tilde{g}_{mn}^1\sim \sum_{l=1} {\cal O}
\left(\frac{\tilde{r}_h}{r}\right)^{4l} {\cal O}\left(\frac{g_s\widetilde{M}^2}{N}\right)+ {\cal O}\left(g_sN_f\right)+higher\;
order
 \nd
 Thus at zeroth order in our perturbation, (\ref{F2sol}) is
 an exact solution of (\ref{dF2}) and (\ref{F2}). Although the world volume flux is evaluated at zeroth order in our
 perturbative parameter, the flux $\widetilde{F}_2$ gives rise to linear order terms ${\cal O}\left(\frac{g_s{\cal M}^2}{N}\right)$  
 in the supergravity action and eventually regularizes it. This is not surprising since the background magnetic field $B_2\sim {\cal O}
 \left(g_sM\right)$ in the KT solution also gives rise to linear order terms ${\cal O}\left(\frac{g_sM^2}{N}\right)$ in the supergravity
 action.    
   
    On the other hand, using the form of the flux, one readily observes that the
 action $S_{I},S_{II}$ are now only a functional of $\tilde{\phi}_1(r)$  and $\tilde{\phi}_2(r)$ respectively. Variation of the action with 
 respect to $\tilde{\phi}_i(r)$
 gives the following equations
 
 \bg\label{embedding-BH}
 &&\frac{e^{\phi}\; r^5\; \tilde{\Phi}_i'}{\sqrt{1+\frac{r^2}{6}\tilde{\Phi}_i'^2}}={\cal C}_i\nonumber\\
 &&\tilde{\phi}_i=\int dr\; e^{-B}\;\frac{\partial \widetilde{\Phi}_i}{\partial r} 
 \nd   
 where $i=1,2$ and ${\cal C}_i$ are constants. The above equation has a simple solution \footnote{
  We are essentially ignoring the running 
 of the
 dilaton on the world volume of the brane. Note that the dilaton field sourced by Branch I  on Branch I is zero but the field sourced by
 Branch II is finite on Branch I. Here we are neglecting the running of the field on Branch I sourced by Branch II and 
 vice versa. If we account for the running, the brane profile will get ${\cal O}(g_sN_f)$ corrections
  which can be ignored since we only consider the profile at
 zeroth order.}
 \bg
 {\rm cos}\left(\frac{4}{\sqrt{6}}\tilde{\Phi}_i\right)=\frac{r_0^4}{r^4}
 \nd
where $r_0$ is a constant. 
\begin{figure}[htb]\label{gauge}
       \begin{center}
\includegraphics[height=6cm]{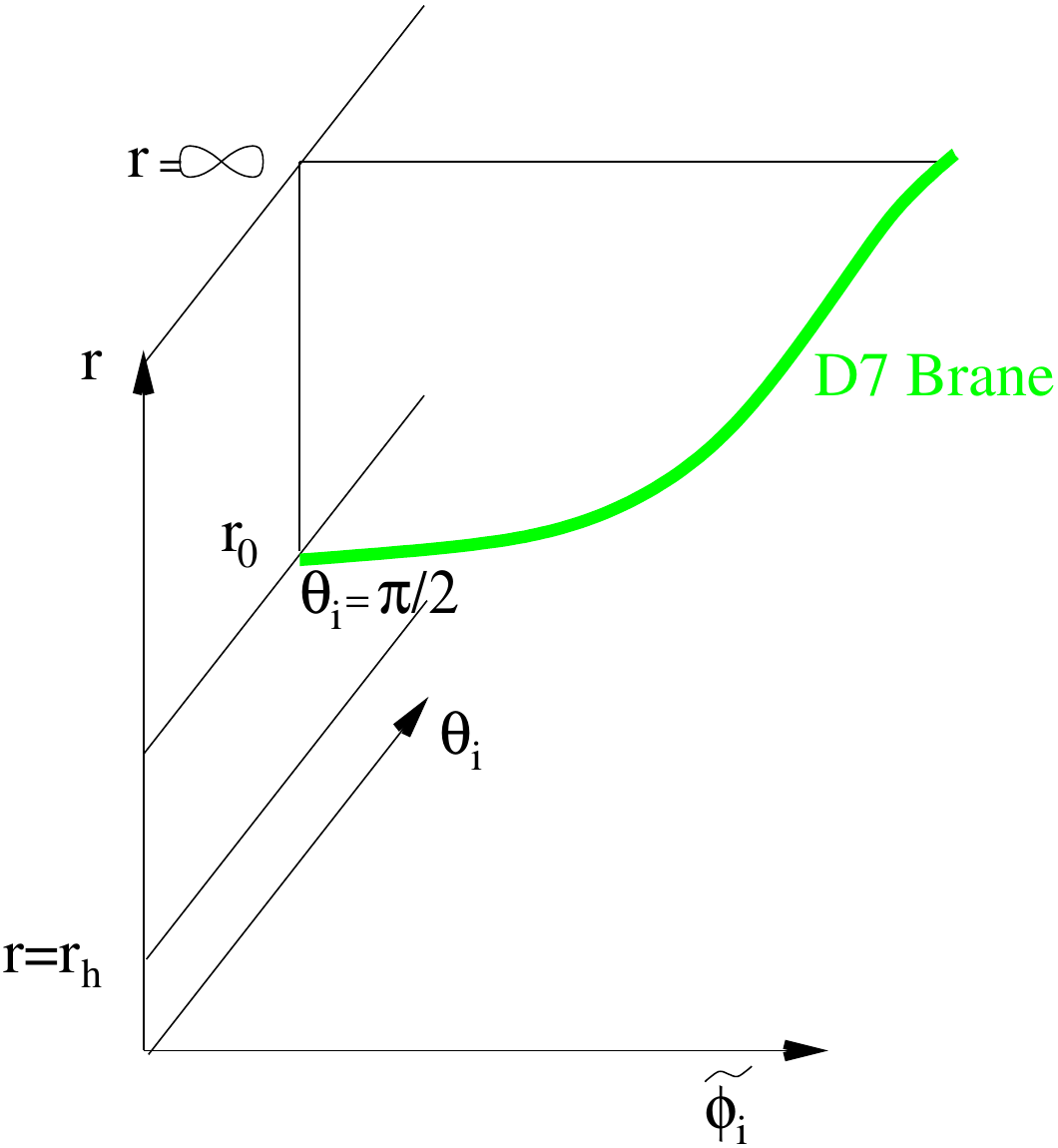}
        \caption{D7 brane embedding; $i=1$ or $2$ corresponds to Branch I or II. }
       \end{center}
        \end{figure} 
Note that $\tilde{\Phi}_i(r_0)=0$ and $\tilde{\Phi}_i(\infty)=\pm\frac{\sqrt{6}\pi}{8}$. However, we pick 
$\tilde{\Phi}_i(\infty)=+\frac{\sqrt{6}\pi}{8}$ for both branches I and II of the D7 branes.   
We can invert $\tilde{\Phi}_i(r)$ to obtain $r(\tilde{\Phi}_i)$ and observe that $r$ has a minimum
given by $r_0$. Thus both branches of D7 brane extend from $r=\infty$ to $r=r_0$. For $r_0=0$, we have $\tilde{\phi}_i=constant$, and  the D7 brane
stretches from the tip $r=0$ to $r=\infty$. Observe that each branch is similar to the embedding considered in 
\cite{Kuperstein:2008cq}. The stability of such embeddings were studied in \cite{Dymarsky:2010ci} and thus we expect our
embedding to be stable under perturbations. A heuristic argument for stability could be the observation that the $D7$ world
volume (\ref{SI}) and (\ref{SII}) are  independent of warp factors at linear order in ${\cal O}(\epsilon)$ and thus the brane
sees a flat geometry. Hence there is no gravitational pull and the embedding is stable.

Note that the $D7$ brane embedding (\ref{embedding-BH}) is well defined for $r_h<r_0$ with $r_h$ being the horizon i.e. 
$e^B(r_h)=0$. When $r_0=r_h$, (\ref{embedding-BH}) implies that $\tilde{\phi}_i(r_h)$ is divergent and thus the embedding is not well
defined.  Thus we will first
consider the localized sources to be outside of the horizon. Then whether we have a black hole
or not,  the boundary values are $\tilde{\phi}(r_0)=0$ and $\tilde{\phi}(\infty)=+\frac{\sqrt{6}\pi}{8}$ for both branches I and II of the D7 branes. However, 
 the shape of the embedding are quite different in the presence of the black hole. Again, by inverting, we can write
 $r(\tilde{\phi})$ and conclude
 that the D7 brane ends at $r=r_0$. 
 
 As we consider larger horizons, we will have $r_h=r_0$ and then the D7 brane will simply fall into the black hole and 
 there will be no action for the brane. Note that the supergravity background without any localized sources  already has non-zero
 three form flux $G_3^0$, sourced by the $D5$ branes of the dual gauge theory. So even at the absence of any localized sources in
 (\ref{Action}), we consider a charged black hole. The addition of localized source for small black holes (that is $r_h<r_0$) 
 then alters $G_3^0\rightarrow G_3$ by inducing additional charges and modifies the total charge
 of the black hole. When $r_h\ge r_0$, the localized sources fall into the black
 hole and we do not have an action for them. However, we must account for their induced charges and
  consider $G_3\neq G_3^0$ even without
 any localized sources outside the horizon. We can choose boundary conditions such that the induced charge is negative and exactly cancels the
 background charge. Thus when $r_h\ge r_0$, the induced charge fall into the black hole and we can simply consider $G_3=0$
 outside the horizon. We will elaborate these points in the following sections.

  Now for the case $r_h<r_0$, that is the brane is outside the horizon and acts as a localized source for
 the world volume fluxes $\widetilde{F}_2$, we can solve the two equations in (\ref{G3}). The solution is 
 \bg \label{G3solu1}
F_3&=&\frac{M\alpha'}{2}\omega_3+4\kappa_{10}^2{\cal M}N_f\alpha' \mu_7 \Bigg(F(r) \widetilde{\omega_3}^1+H(r)\widetilde{\omega_3}^2\Bigg)\nonumber\\
  H_3&=&  \frac{\ast_6\left(e^{B}F_3\right)}{{\rm Im}\tau} 
 \nd 
where $\ast_6$ is the Hodge star for the metric $g_{mn}$ and we have neglected second and higher order in ${\cal O}(g_s)$. Also we have defined the three forms 
$\omega_3,\widetilde{\omega_3}^1,\widetilde{\omega_3}^2$ and the scalar functions $F(r),H(r)$ as follows 
\bg\label{forms}
\omega_3&=&g^5\wedge \left(g^1\wedge g^2+g^3\wedge g^4\right)\nonumber\\
\widetilde{\omega_3}^1&=&\Gamma_1 \left(d\psi +a \;d\phi_2\right)\wedge
 \Omega_1
 -\Gamma_2 \left(d\psi +a\;d\phi_1 \right)\wedge
 \Omega_2\nonumber\\
 \widetilde{\omega_3}^2&=&\frac{{\rm Im\tau}}{{\cal M} }\bigg[\Gamma_1 {\rm cos}(\theta_1) F^I_{\phi_2\theta_2}  d\phi_1\wedge d\phi_2\wedge d\theta_2 
 -\Gamma_2 {\rm cos}(\theta_2)
 F^{II}_{\phi_2\theta_2} d\phi_1\wedge d\phi_2 \wedge d\theta_1\nonumber\\
 &+&d\psi \wedge \big(\Gamma_1 {\rm cos}(\theta_1)F^I_{\psi\theta_2} d\theta_2\wedge d\phi_1-
 \Gamma_2 {\rm cos}(\theta_2)F^{II}_{\psi\theta_1}d\phi_2\wedge d\theta_1\big)\bigg]\nonumber\\
H(r)&=&1\;\;{\rm for}\; r\ge r_0 \nonumber\\
&=& 0 \;\;{\rm for}\;  r<r_0\nonumber\\
F(r)&=&\int^r du  \frac{\sqrt{1+\frac{u^2e^{2B(u)}}{6}\left(\frac{d\phi_1}{du}\right)^2} H(u)}{ue^{B(u)}}
\nd

The fluxes in (\ref{G3solu1}) are the key results of our analysis. They represent the total RR and NS-NS three form flux in the
presence of world volume fluxes on D7 branes. For small black holes that is $r_h<r_0$, (\ref{G3solu1}) gives the fluxes in the
presence of a black hole while by setting $\tilde{r}_h=0$, one obtains the fluxes in vacuum.

Now using the internal metric as $\widetilde{g}_{mn}^0$ with $g_{mn}=e^{-2A-2B}\widetilde{g}_{mn}^0$ and 
the world volume flux $\widetilde{F}_2$ as in (\ref{F2sol}), one can readily check that 
  indeed the fluxes in 
(\ref{G3solu1}) solves the
equations (\ref{G3}) up to linear order in our perturbation. 
 In the extremal limit, $\tilde{r}_h=0, B=0$ and  one readily gets that
$\ast_6 G_3=iG_3$, that
is $G_3$ is ISD in the absence of a black hole.   

Note that the $G_3$ above is obtained using $\widetilde{F}_2$ that is zeroth order in our perturbative parameter $\epsilon$. 
 Thus the  solution we presented here for $G_3,\widetilde{F}_2$ are exact up to zeroth order in $\epsilon$. For higher order corrections to 
 the flux, we must exactly solve the Einstein equations (\ref{Ricci_Min}) to all order in $\epsilon$ to obtain $\tilde{g}_{mn}^1$ and use the metric to find closed, self dual
$\widetilde{F}_2$. Then we can solve (\ref{G3}) using $\widetilde{F}_2$ at higher order and obtain $G_3$ that will include higher order terms
in ${\cal O}(\epsilon)$.

 Now observe that
 the integrand appearing in the definition of $F(r)$ is only non-vanishing for $r_0 \le r$ and thus we can choose boundary
 conditions such that $F(r)=0$ for $r\le r_0$.  
 Then we can choose ${\cal M}$ such that 
$\lim_{r\rightarrow \infty} \int F_3\rightarrow 0$ and eliminate the UV divergences of Klebanov-Strassler theory as shown in the
following section. 

\subsection{Brane engineering: From non-conformal confining IR to conformal UV}
The supergravity solution along with fluxes presented in the previous section may arise from a specific configuration of branes
placed in conifold geometries. Before going into the exact set up of branes, we briefly review two related configurations of
branes that give rise to Klebanov-Witten and Klebanov-Tseytlin/Klebanov-Strassler model: 
Place N D3 branes  at the tip of a regular cone [See Fig
2(a)]. At zero temperature, the gauge group is $SU(N)\times SU(N)$
with bi-fundamental fields $A_i,B_j, i,j=1,2$. This gauge theory has
a conformal fixed plane and the number of D3 branes remains the same
at all energy scales. This is the Klebanov-Witten model \cite{Klebanov:1998hh}. 

Now if we put another stack of D5 branes that
wraps the vanishing two cycle at the tip of the cone [See Fig
2(b)] the gauge
theory becomes $SU(N+M)\times SU(N)$ with the bi-fundamental fields,
and it is no longer conformal. The $SU(M+N)$ sector has $2N$
effective flavors while the $SU(N)$ sector has $2(N+M)$ effective
flavors thus it is dual to the $SU(N-M)\times SU(N)$ gauge theory
under an Seiberg duality. Under a series of such dualities which is
called cascading, at the far IR region the gauge theory can be
described by $SU(M)\times SU(K)$ group, where $N=lM+K$, $l$, $0\leq
K<M$ are positive integers. Now the number of `actual' D3 branes N
is no longer the relevant quantity, rather $N\pm pM$
where $p$ is an integer describes the D3 brane charge. We take $K=0$ in all our analysis, so at
the bottom of the cascade,
 we are left with  ${\cal N}=1$ SUSY $SU(M)$ strongly coupled
 gauge theory which looks very much like strongly coupled SUSY QCD.
\begin{figure}[htb]
                \begin{center}
		\vspace{- 0.5 cm}
		\centering
		\subfigure[]{\includegraphics[width=0.50\textwidth]{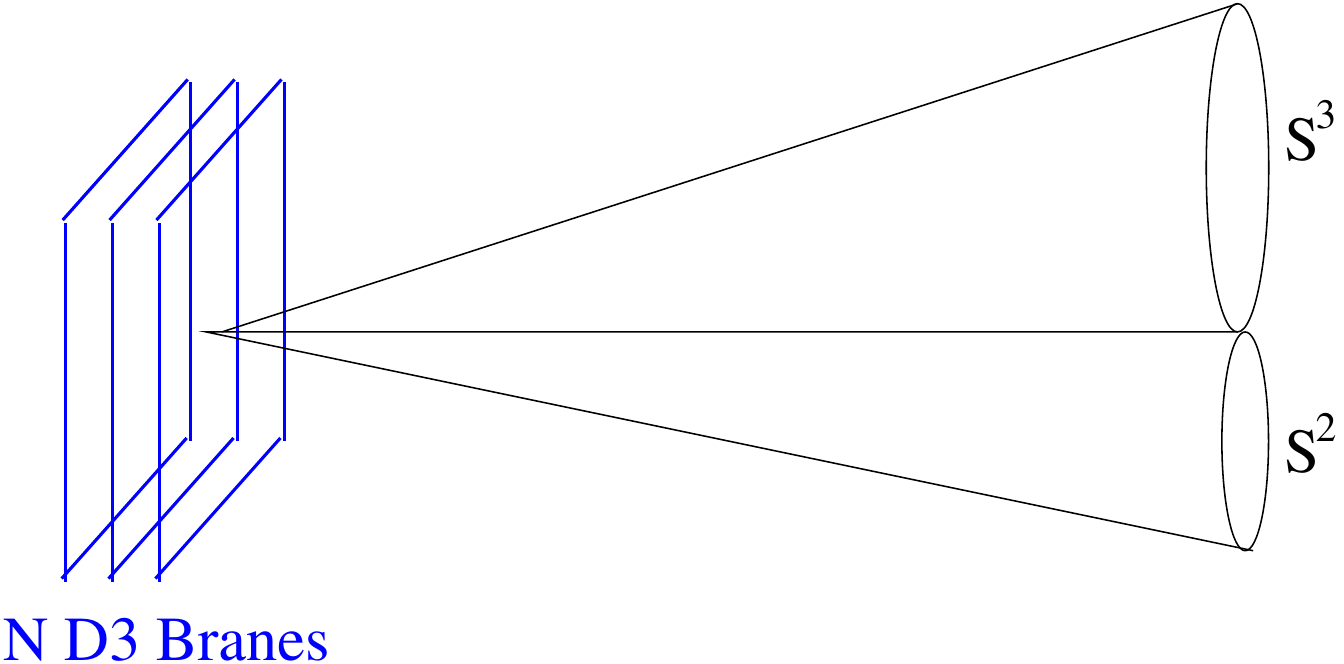}}
		\subfigure[]{\includegraphics[width=0.5\textwidth]{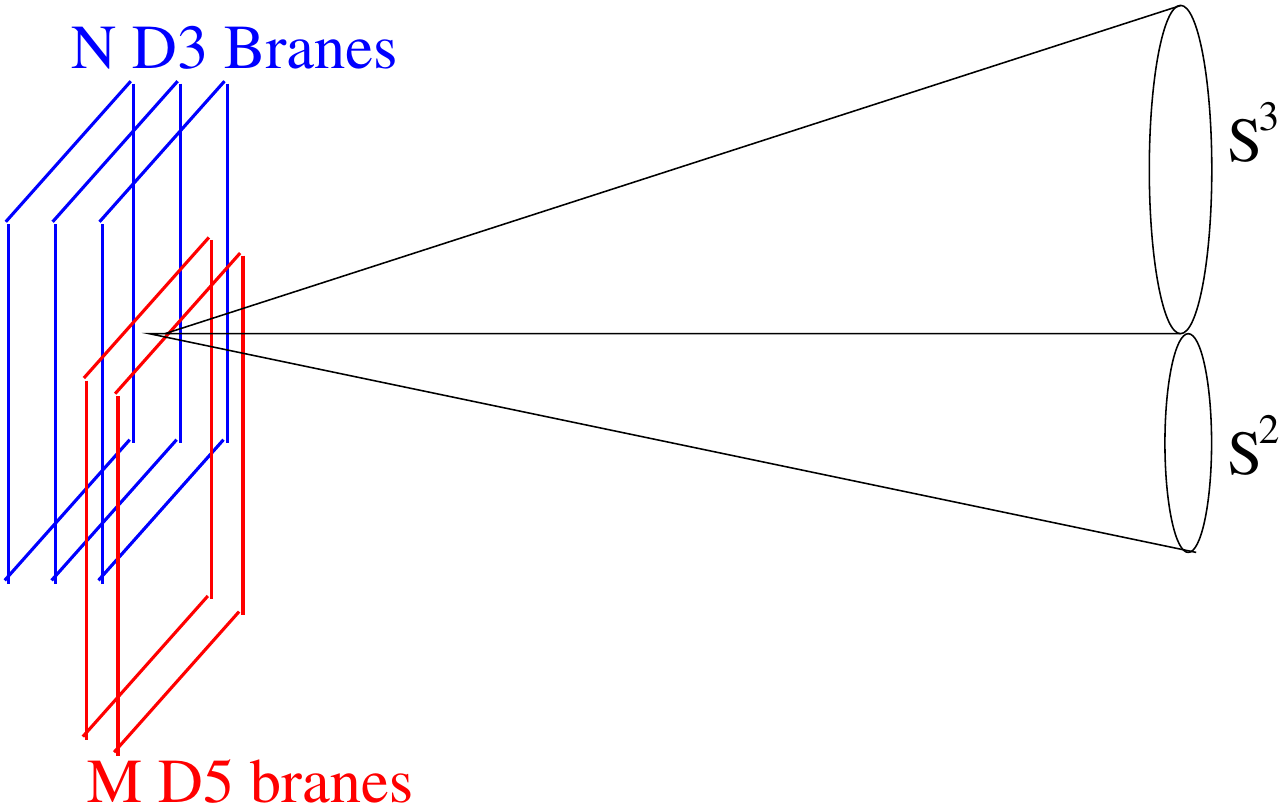}}
        
		\caption{{Brane construction of (a) conformal field theory and (b) non-conformal field theory without UV completion. }}
\end{center}
\end{figure}

 Due to the strong coupling at the IR, the gauge theory develops non-perturbative superpotential \cite{Affleck:1983mk} and  breaks the
 $Z_{2M}$ R symmetry  down to $Z_2$ group. Since the complex fields $A_i,B_j, i,j=1,2$ also describe the complex coordinates of the cone,
 the breaking of the R symmetry  modifies the geometry from a regular to a deformed cone \cite{KS}. Thus to capture 
 the IR modification
 of the gauge theory, we must consider the {\it warped deformed cone}. The warped deformed cone dual to the confining gauge
 theory was proposed by Klebanov-Strassler, while the large `r' region of the geometry was studied by Klebanov-Tseytlin. 
  \begin{figure}[htb]\label{gauge}
       \begin{center}
\includegraphics[height=8cm]{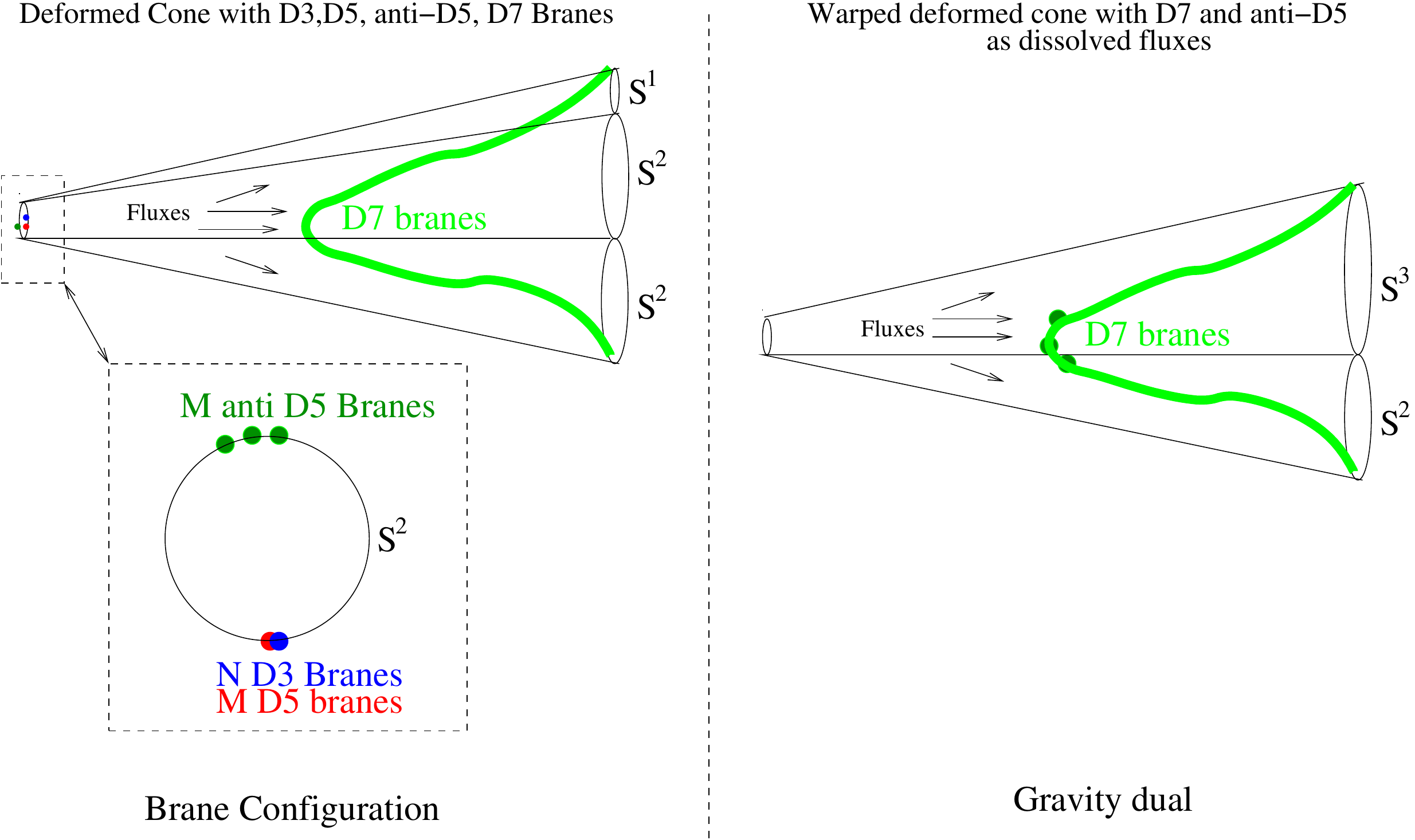}
        \caption{Brane configuration and the dual gravity in the extremal limit for a UV regular theory.}
       \end{center}
        \end{figure} 
 Now observe that the gauge group becomes $SU(M)$ only at the far IR
while at high energies, it can be described by
$SU(k(\Lambda)M)\times SU((k(\Lambda)-1)M)$ group, with $k(\Lambda)$
increasing with energy. Thus the UV of the gauge theory has
divergent effective degrees of freedom and looks nothing like QCD.
Although the confined phase of the gauge theory may resembles
${\cal N}=1$ SUSY QCD, the
deconfined phase of the gauge theory is quite different.
 
  Note that this UV divergence resulting from the Seiberg duality cascade is solely due to the presence of $D5$ brane charge. On
  the other hand, confinement is also a result of the $D5$ branes. To obtain a gauge theory that confines in the IR but
   does not have diverging degrees of freedom at the UV, we need to annihilate the effect $D5$ branes at high energies while
   keeping the theory unchanged at the IR. 
   
   This can be done by adding $M$ anti-five
branes separated from each other and  from the $D3/D5$ branes at the tip of the cone. To obtain this separation, we must blow up one of the
$S^2$'s at the tip and give it a finite size - which essentially means putting a resolution parameter. 
 The brane setup is sketched in the left figure of Fig. 3. 
 The separation gives masses $\Lambda_0$ to the $D5/\bar{D}5$ strings and at scales less than the mass, the
 gauge group is $SU(N+M)\times SU(N)\times U(1)^M$ where the additional $U(1)$ groups arise due to the massless strings ending on the same 
 $\bar{D}5$
 brane.  At scales much larger than $\Lambda_0$, $D5/\bar{D}5$ strings are excited and we have $SU(N+M)\times SU(N+M)$
 gauge theory. Essentially $M$ pairs of $D5/\bar{D}5$ branes with fluxes are equivalent to $M$ number of  $D3$ branes and hence
 they contribute an additional $M$ units of $D3$ charge to Klebanov-Witten theory, resulting in $SU(N+M)\times SU(N+M)$.
     For $\Lambda<\Lambda_0$, i.e. at  low energy, gauge theory is best understood as arising  from the set up of Fig. 2(b) (since the
 modes from 
$\bar{D}5$ branes are not excited), while at high energy $\Lambda\gg \Lambda_0$, the gauge theory is best described as arising from  
left figure of Fig. 3.

Now of course the presence of anti branes will create tachyonic modes and system will be unstable. To stabilize the system, we
need to add world volume fluxes on the $D5/\bar{D}5$ branes. Alternatively, we can introduce $D7$ branes and absorb the anti 
D5
branes as gauge fluxes on the D7 branes. Then a stable configuration of $D7$ branes with gauge fluxes in the presence of
coincident $D3/D5$ branes will be equivalent to
 stable configuration of  coincident $D3/D5$ branes and anti-D5 along with $D7$
branes.  

In the dual geometry, these $D7$ branes appear as stable embeddings in {\it warped} deformed conifold geometry with world
volume fluxes. The embedding $\tilde{\phi}_i(r)$ presented in the previous section along with the world volume flux
$\widetilde{F}_2$ are precisely the values that give a stable configuration. Thus the gravity solution presented in the
previous section describe a gauge theory that arises from anti-$D5$ branes absorbed in $D7$ branes with coincident $D3/D5$
branes at the tip of a regular cone . In the next section we will see how different regions of the dual
geometry represent different regimes of the gauge theory, from the UV modes to the IR modes. In Fig 3 we sketched the brane configuration
and the dual geometry.

In fact, with the solution for the flux presented in the previous section,  the total effective $D5$ charge
 in the gauge theory can be  obtained 
for $r_h<r_0$: 
   \bg
   M_{\rm eff}^{\rm total}(r)=\frac{1}{4\kappa_{10}^2\mu_5}\int F_3=\frac{1}{4\kappa_{10}^2\mu_5}\int \left[\frac{M\alpha'}{2}\omega_3
   +4\kappa_{10}^2{\cal M}N_f\alpha' \mu_7 F(r)
   \widetilde{\omega_3}^1\right]\nonumber\\
   \nd
    This includes the background charge $M$ arsing from $D5$ branes at the tip of the regular cone and the induced $D5$ charges
     due to the presence of the world volume flux $\widetilde{F}_2$ on the $D7$ branes. The latter is given by 
     \bg\label{Mind}
     M_{\rm eff}^{\rm induced}(r)&=&\frac{{\cal M}N_f\alpha' \mu_7 }{\mu_5}\int F(r)
     \widetilde{\omega_3}^1\nonumber\\
     \nd
     From the form of $F(r)$ given in (\ref{forms}), one obtains that $| M_{\rm eff}^{\rm induced}(r)|$ increases with $r$ and
     $ M_{\rm eff}^{\rm induced}(r)=0$ for $r\le r_0$. When we consider the Gaussian surface integral of (\ref{Mind}) 
     for $r>r_0$,
     we enclose the
     induced charges and larger  $r$ encloses more induced charge. We can choose the sign of the induced charge to be negative
     and then decrease total $D5$ charge for increasing $r$. In fact we can choose 
      ${\cal M}, a$ such that       
    \bg \label{Mcond}
     M_{\rm eff}^{\rm induced}(r\rightarrow \infty)&=&-\frac{M\alpha'}{8\kappa_{10}^2\mu_5}\int \omega_3
     \nd 
     This automatically leads to
      \bg
      M_{\rm eff}^{\rm total}(r\rightarrow \infty)=0
      \nd On the other hand, we also demand  
      \bg\label{Bcond}
      \lim_{r\rightarrow \infty}\int B_2\rightarrow 0 
      \nd
      The above condition guarantees that at far UV, the gauge couplings $(g_1,g_2)$ of $SU(N+M)\times SU(N+M)$ theory
       become identical and we essentially have two copies of the same gauge group
       $SU(N+M)$\footnote{Since $\frac{1}{g_1^2}-\frac{1}{g_2^2}\sim \int B_2$, when $\int B_2\rightarrow 0$, $g_1=g_2$ }.
       
      The conditions (\ref{Mcond}), (\ref{Bcond}) gives the value of the constants 
     ${\cal M}$
     and $a$. Observe that (\ref{Mcond}) can be satisfied with any value of $a$ by considering  orientation of the Gaussian
     surface. In particular, the surface $d\phi_1\wedge d\theta_1$ has opposite orientation to the surface $d\phi_2\wedge
     d\theta_2$. Then using the symmetry between $\theta_1$ and $\theta_2$, one readily obtains that the constant $a$ is not
     determined by (\ref{Mcond}) and we get 
      \bg \label{calM}
      {\cal M}\sim -\frac{M}{\lim_{r\rightarrow \infty}F(r)}
      \nd
      since $\int \widetilde{\omega_3}^1$ is finite.
      This also means that $\widetilde{M}$ in our perturbative parameter (\ref{ppara}) is of ${\cal O}(M)$. With the explicit form
      of $B_2$ given in the appendix, we can use (\ref{Bcond}) to determine $a$.

     For the dual gauge theory, (\ref{Mind}) along with (\ref{calM})  implies  that for energy  scale 
     $\Lambda >\Lambda_0\sim r_0$, we have induced anti-D5 branes with charge $ M_{\rm eff}^{\rm induced}<0$ and the total
      effective $D5$ charge at scale $\Lambda\rightarrow \infty$ has vanished. Thus at the far UV, we are only left with $D3$
      charges and Seiberg cascade has terminated. Since we also have $D7$ brane, the axio-dilaton field $\tau$ will be running
      according to (\ref{Imtau})
      and the gauge coupling will also run. But the running of the gauge coupling ($g_{\rm YM}^2$) is of ${\cal O}(g_s^2)$ and can be neglected. Alternatively,
       by considering additional assembly of $D7/{\bar D}7$ branes  the
      as discussed in detail
      in  \cite{Mia:2010tc}\cite{Mia:2011iv} \cite{fangmia} \cite{Chen:2012me}, we can effectively make the axio-dilaton field
      constant.  Then at the far UV ($\Lambda\gg \Lambda_0$), we end up with a gauge theory with color symmetry $SU(N+M)\times SU(N+M)$, where gauge
      couplings $(g_1,g_2)$ are identical and do not run. At the far IR ($\Lambda\ll \Lambda_0$), 
      the gauge theory cascades down to a single group $SU(M)$ and confines. This way we obtain a gauge theory that is UV
      conformal and IR confining- which are common features with QCD. In the next section, we will explore the thermodynamics
      of the gauge theory and make direct connections to QCD.

Coming back to the flux analysis, since $\int_{r\rightarrow \infty} F_3=0,\int_{r\rightarrow \infty} B_2=0$, we readily get 
\bg
\int G_3\wedge \ast_6 \bar{G}_3
\nd      
is negligible in the large $r$ region.
Thus the bulk action (\ref{Action}) gets negligible contribution from $G_3$ for $r\gg r_{0}$.
 By integrating (\ref{Action}) over the angular coordinates $\psi,\phi_i,\theta_i$, we can obtain an action that describes 
 $AdS_5$
 geometry for $r\gg r_{0}$. Hence, addition of the localized sources has allowed us to patch together a deformed warped cone at
 small $r$ to an $AdS_5\times T^{1,1}$ like geometry at large $r$.

 Finally, observe that we consider the induced charge (\ref{Mind}) when the source is outside the horizon, that is we still
 have small black holes $r_h<r_0$. When $r_h\ge r_0$, the D7 brane falls into the black hole, along with the effective
 anti-D5 charge. These anti-D5 charges will neutralize the D5 charge and thus for large black holes $r_h\ge r_0$, we can simply
 consider $G_3=0$ outside the horizon. Then the only non-trivial flux in (\ref{Action}) would be $\widetilde{F}_5$, as will be
 discussed in detail in section 3.2.

 \section{Thermodynamics of the gauge theory}
Using the gravity solution of the previous section, we can obtain the partition function of the gauge theory using the identification
\begin{eqnarray}\label{KS16}
\mathcal{Z}_{\rm gauge}&=&e^{-F/T}=\mathcal{Z}_{\rm gravity}\simeq e^{-S^{\rm ren}_{\rm gravity}}\nonumber\\
S^{\rm ren}_{\rm gravity}&=&S_{\rm bulk} +S_{GH}+ S_{\rm counter}
\end{eqnarray}
where $S_{\rm bulk}=S_{\rm SUGRA}+ S_{\rm loc}$, $S_{\rm loc}$ is the action for the localized sources (branes,
anti-branes, or other localized manifolds), $S_{GH}$ is the Gibbons-Hawking boundary term and $S_{\rm counter}$ is counter
terms required to regularize the action. Then using thermodynamic identities we can directly obtain free energy, pressure and entropy from the
partition function.
 
 However, the action $S_{\rm total}$ gives rise to geometries with or without a black hole and we denote them by $X^2$ and $X^1$. The on shell
 values of the action including the Gibbons-Hawking terms for the two geometries $X^1$ and $X^2$ are distinct and we denote them by $S^1$ and $S^2$. The temperature of the dual
 gauge theory corresponding
 to the manifold $X^1,X^2$ can be obtained by identifying the periodicity of Euclidean time. However, for a manifold with singularity, this
 period is fixed by the nature of the singularity, while for a regular manifold, the period is arbitrary. For a given temperature of the dual
 gauge theory,
 the geometry with less on-shell value for the action will be preferred. Using self-duality of five form flux, and ignoring second order terms
 in ${\cal O}(\epsilon)$ that arise from the axio-dilaton field, we readily obtain the {\it bulk Euclidean} on-shell action:
 \bg \label{S} S_{\rm
bulk}&=&\frac{1}{2\kappa_{10}^2}\Bigg[\int
d^8x\int_0^{\beta}d\widetilde{\tau} \int_{r_h}^\infty
dr\sqrt{G_2}\left(-\frac{G_3\cdot \bar{G}_3}{24{\rm
Im}\tau}\right)\nonumber\\
&-&i \int \frac{C_4\wedge G_3\wedge \bar{G}_3}{4 {\rm
Im}\tau }\Bigg]+S_{\rm loc}\nonumber\\
S_{\rm loc}&=& N_f\;S_{Dp}
\nd
 The above action is obtained from the on-shell value of (\ref{Action},\ref{Action1}) after wick rotation $t=i\widetilde{\tau}$ 
 and imposing periodicity
 $\beta$. In particular we used $S^{\rm E}=-iS^{\rm M}$, where $S^{\rm M}$ is the action using Minkowski metric and then 
 wick rotating the on-shell
 value and $S^{\rm E}$ is the Euclidean action. 
 
 In order to study the gauge theory at different energy scales, it is useful to identify two regions in the dual geometry 
 according to the radial distance $r$. This is particularly instructive since gravitons coming from the  small $r$ region are red shifted 
 compared to large $r$ region as measured by an observer at a fixed $r=r_c$. Thus, small $r$ region is dual to low energy modes  while large
 $r$ region accounts for the high energy modes of the gauge theory. 
 
 Before discussing these two regions in detail, we
 would like to point out the connection between our gravitational description and 
  Wilsonian Renormalization Group (RG) flow. The RG flow of the gauge couplings $(g_1,g_2)$ can be obtained
  from the dual flux $B_2$ and the dilaton field $e^{\phi}$ from the following relation:
  \bg\label{RGflow}
  \frac{1}{g_1^2}+\frac{1}{g_2^2}&=&e^{-\phi}\nonumber\\
  \frac{1}{g_1^2}-\frac{1}{g_2^2}&=&e^{-\phi}\int_{S^2} B_2
  \nd
  
  By replacing the radial coordinate with energy scale i.e. $r\rightarrow \Lambda$, and using the expression for $B_2, e^{\phi}$ as
  given in   (\ref{B2}),(\ref{Imtau}) in 
  extremal limit ($r_h=0$), one readily obtains the running of the gauge couplings $(g_1(\Lambda),g_2(\Lambda))$ 
  with scale $\Lambda$. The flux $B_2$ and dilaton field $e^{\phi}$ was obtained using the bulk action (\ref{Action}), 
  (\ref{Action1}) which
  describes the entire geometry, from $\rho=0$ to $r=\infty$. However, if we consider the action by restricting the radial integral
  up to $r=r_0$ and we neglect the localized sources, then the action (denoted by $S_{\rm R_1}$) describes the 
  fluxes in small $r$ region, i.e.
  $r<r_0$. We denote this small $r$ region $r<r_0$ as region 1. In particular, the action $S_{\rm R_1}$ and 
  $S_{\rm SUGRA},S_{D7}$ in (\ref{Action}), 
  (\ref{Action1}) both give identical result for flux and the dilaton field in region 1, up to linear order in ${\cal O}(\epsilon)$.
  Thus whether we use $S_{\rm R_1}$ or $S_{\rm SUGRA}+S_{D7}$ as the action, both will describe identical RG flow according to
  (\ref{RGflow}). 
  Hence $S_{\rm R_1}$ can be thought of as the Wilsonian effective action that describes the IR of the gauge theory. 
  
  However, we need
  to be careful in carrying out this analogy. The on-shell values of $S_{\rm R_1}$ and $S_{\rm SUGRA}+S_{D7}$ are distinct and so
  are the corresponding partition functions. On the other hand, in Wilsonian RG flow, the partition function is independent of the
  flow. Thus the `flow' in going from $S_{\rm SUGRA}+S_{D7}$ to $S_{\rm R_1}$ is not identical to the Wilsonian flow, even though
  the beta functions follow the analogy.

   In the following we first discuss region 1 and then the inclusion
 of large $r$ region, denoted by region 2.
   
  \subsection{Region 1, $r<r_0$}
  
  \begin{figure}[htb]\label{gauge}
       \begin{center}
\includegraphics[height=6cm]{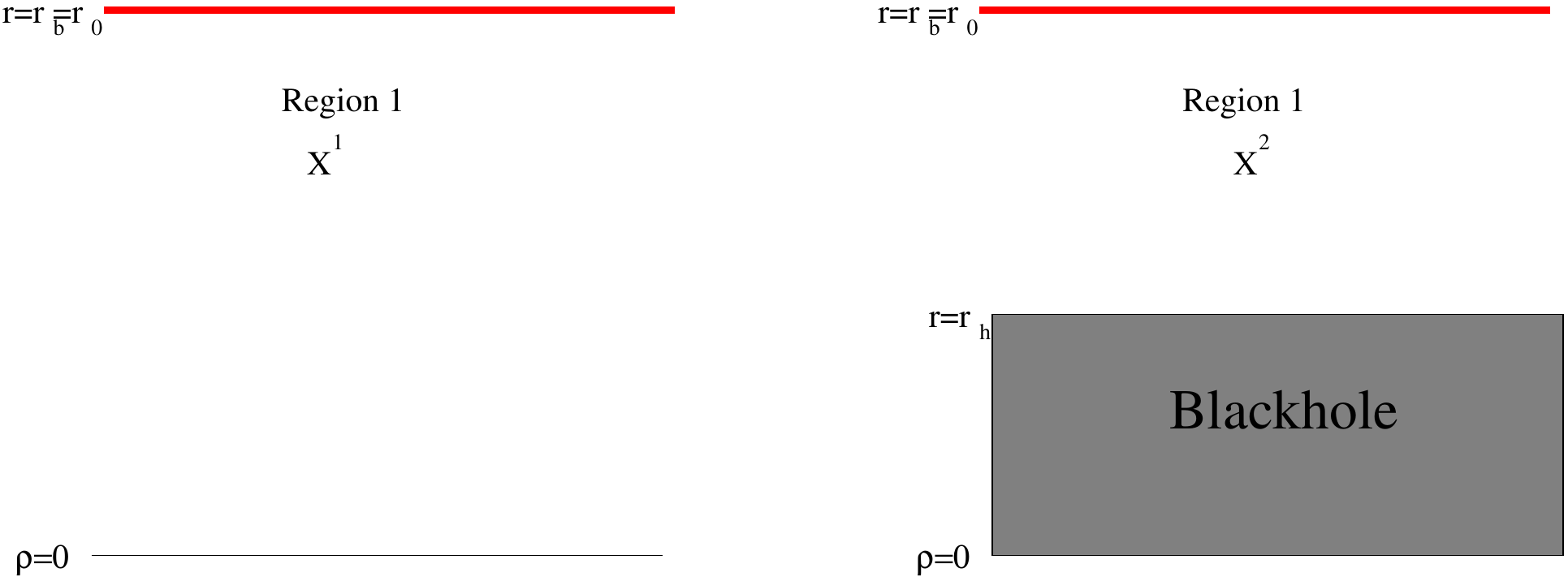}
        \caption{Region 1 with boundary $r_b=r_0$, with or without a black hole.}
       \end{center}
        \end{figure} 
   First consider this region for extremal manifold $X^1$. The manifold is regular and 
  the metric is given by (\ref{metric}) with the internal metric
  (\ref{inmate}) in the limit $B=0$. The warp factor $h(\rho)\equiv e^{-4A}$ has no singularities in this region and thus we can impose any
  periodicity $\beta$ of Euclidean time $\widetilde{\tau}=-it$, after wick rotation. Thus, region 1 can describe {\it any} temperature $T_1\equiv
  \beta_1^{-1}$. Furthermore, this
  region does not include the localized sources and in the extremal limit $B=0$,  three form flux is ISD: $\ast_{6}G_3=i G_3$
  with $C_4=e^{4A}d\widetilde{\tau}\wedge dx\wedge dy\wedge dz$. 
  Then for geometry $X^1$ in region 1, we have $G_3\wedge\ast_{10} \bar{G}_3=-iC_4\wedge G_3\wedge \bar{G}_3$.  Thus if we
  restrict the integral in (\ref{S}) to region 1, we get 
  \bg
  S^1_{\rm R_1}&=&\frac{1}{2\kappa_{10}^2}\Bigg[\int
d^8x\int_0^{\beta}d\widetilde{\tau} \int_{\rho=0}^{r=r_0}
dr\sqrt{G_2}\left(-\frac{G_3\cdot \bar{G}_3}{24{\rm
Im}\tau}\right)\nonumber\\
&-&i \int \frac{C_4\wedge G_3\wedge \bar{G}_3}{4 {\rm
Im}\tau }\Bigg]\nonumber\\
&=&0
  \nd

  Next we can consider the non-extremal manifold  $X^2$ in region 1 which has a horizon $r_h\le r_0$. Again, using our 
  explicit solutions for the fluxes given in (\ref{G3solu1}), we can
  evaluate the on-shell action up to linear order in ${\cal O}(\epsilon)$ for region 1 to obtain \cite{Mia:2012ue}

\bg \label{SR1}
S^2_{\rm R_1}&=&\frac{\beta_2 g_sM^2V_8}{2\kappa_{10}^2N}\int_{r_h}^{r_0}dr\frac{3\tilde{r}_h^4}{16r}\nonumber\\
&=&\frac{3\beta_2 g_sM^2V_8\tilde{r}_h^4}{32\kappa_{10}^2 N}{\rm log}\left(\frac{r_0}{\tilde{r}_h}\right)
\nd

Before analyzing the action, it is crucial to note that the above result is valid {\it only }for a large black holes, 
that is $r_h$ is large. For a small black hole, 
we need to go back to radial coordinate $\rho$ and consider the warp factor at small $\rho$
\bg \label{warp-smallr}
e^{-4A(\rho)}=h(\rho)=c_i\rho^i\nonumber\\
e^{2B(\rho)}=1+\sum_{i=1}b_i\rho^i
\nd
where $c_0\neq 0$, $c_i,b_i\sim {\cal O}(g_sM^2)$ are constants and we have the boundary conditions $h(\rho_h)\neq 0,
e^{2B(\rho_h)}=0$. In the extremal limit when horizon 
$\rho_h=0$, we have $e^{2B(\rho)}=1$ and
$h(\rho)$ is the Klebanov-Strassler warp factor. In the non-extremal limit, we must solve
 (\ref{warp_eq_1}), (\ref{BHfactorA}) and (\ref{GKP_BH}) using the internal metric of the deformed cone $\tilde{g}_{mn}$
 (with $\tilde{g}_{mn}^0$ given by (\ref{inmate}) and $\tilde{g}_{mn}^1$ obtained from solving the second Einstein equation in
 (\ref{Ricci_Min})). In solving these system of coupled equations in small $\rho$ region, 
 the relevant parameter is $\frac{g_sM^2}{N_{\rm eff}(\rho)}$ with

 \bg\label{Neff}
 &&N_{\rm eff}(\rho)=\frac{1}{2 \kappa_{10}^2T_3}\int_{T^{1,1},\rho} \tilde{F}_5\nonumber\\
 &&\int_{T^{1,1},\rho_b} \tilde{F}_5-\int_{T^{1,1},\rho_a} \tilde{F}_5=\int_{\rho_a}^{\rho_b} d\tilde{F}_5
 \nd
 In the above, the surface integral is taken over $\rho=constant$ surface which is a warped $T^{1,1}$. 
 When $\rho_a\rightarrow 0$, warped $T^{1,1}$ shrinks to a three sphere $S^3$ and the surface integral vanishes,
 $\int_{T^{1,1},\rho\rightarrow 0} \tilde{F}_5\rightarrow 0$. Then we get
 \bg
 N_{\rm eff}(\rho)=\int_{0}^{\rho} d\tilde{F}_5
 \nd
 
 Then using (\ref{BI}), one readily gets that $N_{\rm eff}(\rho)$ decreases with $\rho$ and for small enough $\rho$,
 \bg
 \frac{g_sM^2}{N_{\rm eff}(\rho)}>1
 \nd  
 becomes large. Thus {\it for small black holes, we cannot solve the system of equations
 (\ref{warp_eq_1}), (\ref{BHfactorA}),(\ref{GKP_BH}) and the Einstein equations perturbatively, since there is no small
 perturbative parameter like (\ref{ppara})}. Hence, we do not have exact
 expressions for  the on-shell action
 (\ref{S}) when the black hole horizon $\rho_h$ is small. 
 
 However, the region of the geometry near  small horizons is
 crucial in determining the low temperature dynamics of the gauge theory. Especially small black holes will determine the
 thermodynamics of the gauge theory near critical phase transition temperature, which we will see shortly. 
 In fact the small $\rho$ region is dual to QCD like
 $SU(M)$ pure glue theory while large $\rho$ region is dual to a bi-fundamental gauge theory, very different from QCD. Thus,
 the black hole geometry dual to QCD like gauge theory near phase transition cannot be  described by our perturbative approach.
 Although we do not have an exact solution for the metric near horizon for small black holes i.e. $\rho_h$ is small, we can
 find the form of the warp factor $A(\rho)$ and black hole factor $B(\rho)$. Using these forms, we can qualitatively understand
 the behaviors of thermodynamic state functions. We will elaborate  the issue in the following sections, which in fact makes it clear the 
 connection between lattice QCD and our holographic model.  
   
 Only at large $\rho$, (\ref{warp-smallr}) takes the form (\ref{alpha}), with $N\equiv N(r_l)\gg 1$ for some large $r_l$. Then 
 the parameter $\epsilon$
  is very small and our perturbative analysis is valid with the on-shell action given by (\ref{SR1}).
   Since $X^2$ has black hole singularity, the  periodicity $\beta_2$ is not arbitrary and the temperature of the field theory living at surface
  $r_b$ is given by
   \bg \label{Tncft}
T_2(r_b)=\frac{\beta_2^{-1}}{\sqrt{g(r_b)}}\nonumber\\
\beta_2=\frac{4\pi\sqrt{h(r_h)}}{|g'(r_h)|}
\nd
where $g=e^{2B}$ and prime denotes derivative with respect to $r$. 
The two geometries describe the same field theory at the same temperature
on the hyper surface $r=r_b$ if

\bg \label{beta}
T=T_1(r_b)=T_2(r_b)\Rightarrow \beta_1=\beta_2 e^{B(r_b)}
\nd 
Now the Gibbons-Hawking term at boundary $r_b=r_0$ for $X^1,X^2$ is given by \cite{Mia:2012ue}
\bg
S^1_{R_1,GH}&=&\frac{1}{108\kappa_{10}^2}
\Big[4r_0^4+\frac{729g_sM^2}{16 N}r_0^4\Big]\beta_1V_8\nonumber\\
S^2_{R_1,GH}&=&\frac{1}{108\kappa_{10}^2}
\Big[4r_0^4-2\tilde{r}_h^4+\frac{729g_sM^2}{16 N}(r_0^4-(1+d)\tilde{r}_h^4)\Big]\beta_2V_8+\widetilde{\cal I}\nonumber\\
\widetilde{\cal I}&\equiv&{\cal O}\left(\frac{g_sM^2}{N}\right)\sum_{l=1}^{\infty}{\cal
O}\left(\frac{\tilde{r}_h^{4(l+1)}}{r_0^{4l}}\right) 
\nd
Then we get the action difference, 
 \bg \label{delSa}
  \triangle S&=&S^2_{\rm R_1}-S^1_{\rm R_1}+S^2_{R_1,GH}-S^1_{R_1,GH}\nonumber\\
  &=&\frac{3g_sM^2\beta_2 V_8r_h^4}{32\kappa_{10}^2N}
  \left({\rm log}\left(\frac{r_0}{\tilde{r}_h}\right)-\frac{9}{4}-\frac{9}{2} \left[d-\alpha^1\right]\right)+ {\cal I}\nonumber\\
  {\cal I}&\equiv&\widetilde{\cal I}+{\cal O}\left(\frac{g_sM^2}{N}\right)\sum_{l=1}^{\infty}{\cal
O}\left(\frac{\tilde{r}_h^{4(l+1)}}{r_0^{4l}}\right) 
   \nd
where $\alpha^1$ arises from the following expansion  
\bg
e^{2B}=g(r)\equiv 1-\frac{\tilde{r}_h^4}{r^4}\left(1+
 \frac{729 \alpha^1 g_sM^2}{8 N}\right)+ \frac{\tilde{r}_h^8}{r^8} {\cal O}\left(\frac{g_sM^2}{N}\right)+...
\nd

Note that $d$ is determined by the Einstein
equations and the flux equations once boundary conditions are imposed. Both $d$ and ${\cal I}$
 arise from the expansion of $g(r)$ and are sensitive to the near horizon geometry and in particular can be 
obtained from the horizon values of the metric. Furthermore, $d, {\cal I}$ are related to $h(r_h)$, which in turn determines the number of effective 
degrees of freedom at a
temperature $T\simeq r_h/L^2$. Thus from the gauge theory side, $d, {\cal I}$ are both
 related to the effective colors at the thermal scale.  

Without explicitly solving the Einstein equations near
the black hole horizon, we can speculate two boundary conditions by ignoring ${\cal I}$ (which is small 
since $r_0>\tilde{r}_h$ always) : 

\vspace{15pt}

\noindent $\bullet$ $d\le \alpha^1-1/2$: In this case $\triangle S> 0$, which means extremal geometry is favored over black hole. 
In that
 case all black holes in region 1
have higher free energy compared to vacuum and the dual gauge theory is described by extremal geometry $X^1$. 
However note that,
   since there is no black hole horizon in the vacuum geometry $X^1$, the Euclidean renormalized on-shell action $S^{\rm ren}_{\rm gravity}=\beta F$
     is independent of horizon with
    $F$
   independent of $T$. Then using  thermodynamic identity, one  readily gets
  \bg \label{s0}
  s=-\frac{\partial F}{\partial T}=0
  \nd
Thus for $d\le \alpha^1-1/2$ and we ignore ${\cal I}$, region 1 corresponds to {\it confined} phase. 

It is possible that for
$d\le \alpha^1-1/2$ we can still have $\triangle S(\tilde{r}_h=r_h^c)= 0$ if  ${\cal I}(\tilde{r}_h=r_h^c)$ is not small.
Then $r_h^c$ would give the critical horizon. ${\cal I}$ becomes more and more
significant for larger $\tilde{r}_h$, but since we do not have an exact expression for ${\cal I}$, we cannot directly analyze this
case. In fact for a  gauge theory  that behaves similar to QCD near the critical phase transition temperature, 
it is likely that we cannot ignore ${\cal I}$. In particular, we will find out later in this section 
that ignoring ${\cal I}$ results in a conformal
anomaly that only matches the lattice QCD behavior for temperature much larger than $T_c$ and not near $T_c$.

\vspace{15pt}

 \noindent $\bullet$  $d> \alpha^1-1/2$: In this case, using (\ref{delSa}) and ignoring ${\cal I}$, it is possible to obtain $\triangle S=0$ in region 1,
 with the following value for 
 critical horizon
\bg\label{Tc}
r_h^c&=&\frac{r_0}{{\rm  exp}\left(\frac{9}{4}\left[1+2(d-\alpha^1)\right]\right)}
\nd
 Then the 
 corresponding critical temperature is
\bg\label{Tcc}
T_c&=&\frac{r_h^c\left(1+{\cal O}\left(\frac{g_sM^2}{N}\right)\right)}{\pi L^2(r_h^c)}\sim \frac{2r_0}{{\rm 
exp}\left(\frac{9}{4}\left[1+2(d-\alpha^1)\right]\right)\sqrt{27\pi N\alpha'}}\nonumber\\
L^4(r)&\equiv& r^4 h(r)
\nd

We need to keep in mind that the derivation of critical horizon (\ref{Tc}) uses the form of the action (\ref{SR1}), 
which is
only valid for large black holes. But the black hole cannot be too large, since  we are also ignoring ${\cal I}$,
 which is only justified for 
$\tilde{r}_h\ll r_0$. Note that the notion of `largeness' and `smallness' in region 1 is relative to the scale $r_0$: 
If $r_h\ll r_0$, we
have a small black hole while if $r_h\lesssim r_0$, we have a `large' black hole in region 1. 

If $d$ is very large, $r_h^c$ will be small and (\ref{SR1}) will no longer give the exact
on-shell action, since our perturbative analysis will break down. Only for small $d> \alpha^1-1/2$, we will have large enough
$r_h^c$ such that our perturbative analysis holds 
(but in that case ${\cal I}$ can become large and not negligible) and $T_c$ is given by 
(\ref{Tcc}). Each set of value for $(d, {\cal I})$ fixes the geometry and
describes a particular gauge theory, which may or may not resemble QCD near critical temperature.  Small 
(but $d> \alpha^1-1/2$ ) values
of $d$ which makes our perturbation in ${\cal O}\left(\frac{g_sM^2}{N}\right)$ exact, also makes ${\cal I}$ large and thus
computation of $T_c$ less and less exact. Furthermore, with small $d$, the width of the conformal anomaly is small and
distinct from lattice QCD results. This
mismatch with lattice QCD is in fact not unexpected and can be attributed to the fact that we ignored ${\cal I}$. We  will
elaborate the issue in detail in the following section.     

 Coming back to our perturbative gravity analysis, for $T>T_c$, the black hole is preferred and describes the deconfined
phase which has non-zero entropy. On the other hand for $T<T_c$, we have a confined phase with zero entropy described by the extremal geometry.
 Thus at $T=T_c$, there is a first order phase
transition in the gauge theory and we have obtained a gravitational description of it in terms of Hawking-Page phase transition between two
geometries.  

Now to obtain the thermodynamic state function of the gauge theory, we have to obtain the
partition function using (\ref{KS16}) and thus we need the counter terms for $r_0\rightarrow \infty$.  
 The divergent part $\lim_{r_0\rightarrow \infty} \beta_2 r_0^4$ is
dependent on temperature and  thus the counter term will also be dependent on it. Also observe that the regularization scheme adopted in 
\cite{Hawking:1982dh,Witten:1998zw} identifies $\Delta S$ as the regularized gravity action. Hence regularization is not just subtraction of
 the infinite part $\lim_{r_0\rightarrow \infty} \beta_2 r_0^4$, but also a finite part arising from the
vacuum action. Thus, even if $r_0$ is finite, which is the case we are in fact considering, we need counter
terms that depend only on temperature. 

There is an alternative approach to obtain thermodynamic state function. We can use Wald's formula for the gravity Lagrangian
\bg
{\cal L}_{\rm bulk}=R+\frac{\partial_M
\tau\partial^M\bar{\tau}}{2|{\rm
Im}\tau|^2}-\frac{|\widetilde{F}_5|^2}{4\cdot 5!}-\frac{G_3 \cdot \bar{G}_3}{12 {\rm Im}\tau}
\nd
 to compute the
entropy associated with the geometry $X^2$ in region 1. Using the expansion for the warp factors (\ref{alpha}), the form of the internal metric
 (\ref{inmate},\ref{gmn}) and considering only up to linear terms in ${\cal O}(\epsilon)$, we get the Wald entropy
 \cite{wald1}-\cite{wald4}
 \begin{eqnarray}\label{entropyR1}
s_{\rm R_1}&=&\frac{\pi r_h^3 \sqrt{27\pi N}V_8\alpha'}{108\kappa_{10}^2}\left(1+ \frac{a_0 g_sM^2}{N}+\frac{a_1 g_sM^2}{N}{\rm log}\frac{r_h}{r_*}\right)\nonumber\\
V_8&\equiv& V_5\times V_3\nonumber\\ 
V_5&\equiv& \int d\psi d\phi_1 d\phi_2 d\theta_1 d\theta_2 \; \rm{sin}\theta_1 \rm{sin} \theta_2, ~~~ V_3\equiv \int dx\;dy\;dz
\end{eqnarray}
where $a_0,a_1$ are  constants independent of $M,N$ and is determined by  the value
of the warp factor $h$ and internal metric $\widetilde{g}_{mn}$ at the horizon. Using the definition of temperature (\ref{Tncft}) in terms of the horizon, we
can obtain entropy as a function of temperature, 
\bg\label{sfT}
s_{\rm R_1}=\frac{27\pi^6 T^3 N^2V_8}{32\kappa_{10}^2}\left(1+\frac{b_0 g_sM^2}{N}+  \frac{b_1 g_sM^2}{N}{\rm log}\left(T\sqrt{N\alpha'}\right)\right)
\nd
where $b_0,b_1$  are a constants independent of $M,N$. They are
 determined by using the definition of temperature (\ref{Tncft}) and   the horizon value
of the metric.  

Once entropy is known as a function of temperature, we can readily obtain the free energy of the gauge theory dual to region 1, 
\bg\label{FreeER1}
F_{\rm R_1}&&=-\int dT\; s_{\rm R_1}\nonumber\\
&=&-\frac{27\pi^6 T^4 N^2V_8\alpha'^4}{128\kappa_{10}^2}\left(1+\frac{b_0 g_sM^2}{N}-\frac{b_1 g_sM^2}{4N}
+\frac{b_1 g_sM^2}{N}{\rm log}\left(T\sqrt{N\alpha'}\right)\right)\nonumber\\
\nd

 \subsection{Region 1 and region 2, $r>r_0$}      
 \begin{figure}[htb]\label{gauge}
       \begin{center}
\includegraphics[height=6cm]{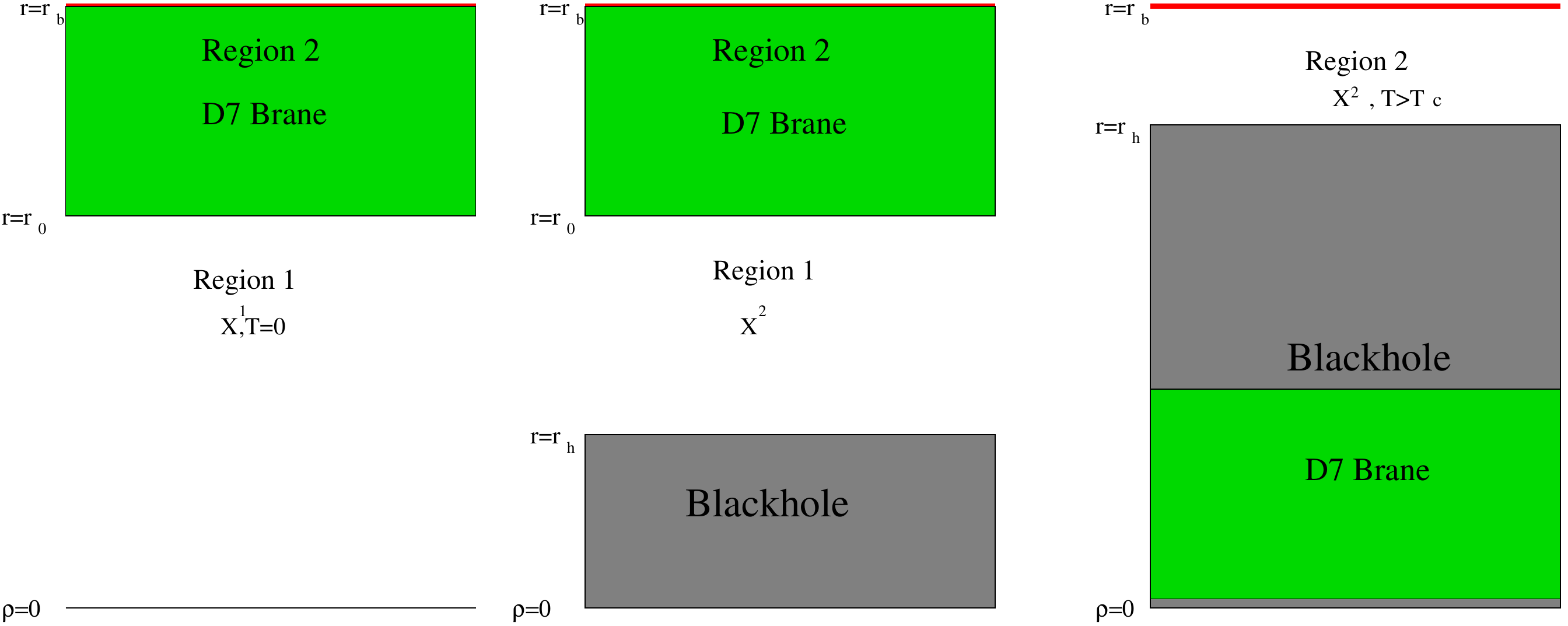}
        \caption{D7 brane in Region 2, with boundary $r_b\rightarrow \infty $, with or without a black hole.}
       \end{center}
        \end{figure} 
  
 We will now consider the on-shell action for both the extremal $X^1$ and non-extremal black hole geometry $X^2$, by including the large $r$
 region of the geometry. Inclusion of localized terms makes $X^1$ a singular manifold, since we are taking the back reaction of
 the local source. Removing the singularity imposes that the period
 $\beta_1\rightarrow \infty$ and thus $X^1$ now {\it only describes zero temperature}. In particular $X^1$ with D7 branes will
 describe a gauge theory with mesons and gluons at zero temperature.

On the other hand, 
 $X^2$ can  have any horizon $r_h$. For $r_h<r_0$, the
  inclusion
 of localized sources gives rise to two singularities to the manifold $X^2$: one at the black hole horizon and the other at the
 location of the brane. We can only remove one of the singularities by imposing a single periodicity of Euclidean time 
 throughout the manifold - 
 but then the metric is not smooth since the other singularity remains. Identifying temperature this way does not result in a
 unique temperature and thus $X^2$ does not describe a gauge
 theory at thermal equilibrium. $X^2$ without localized sources describe gluonic medium at thermal equilibrium and the
 temperature can be identified from the black hole singularity of the manifold. Insertion of D7 brane into the
 geometry is equivalent to immersing  mesons into a gluonic thermal bath. When the system is not at thermal equilibrium,
 mesons and gluons have different temperatures. Thus $X^2$ with D7 branes outside the horizon represents a system that is not at
 thermal equilibrium\footnote{We thank Martin Kruczenski for elaborating this point}. 
 
  Analytic continuation $t\rightarrow i\widetilde{\tau}$ to Euclidean metric and then
 identifying temperature with period of $\widetilde{\tau}$ is applicable for a system at thermal equilibrium. Since $X^2$ including $D7$
 branes do not describe a system at thermal equilibrium, Euclidean action of singular geometry $X^2$ does not represent the free
 energy of the system.   Thus for $r_h<r_0$ and we include 
 region 2, we {\it do not} have multiple description of gauge theory in terms of $X^1,X^2$ and there is no Hawking-Page
 like transition.

For $r_h\ge r_0$, the localized sources will fall into the horizon and now we have a unique definition of temperature as the only
singularity  is the black hole singularity. Thus the black hole geometry with horizon $r_h\ge r_0$ can describe non-zero temperature gauge
theory with the temperature given by the Hawking temperature (\ref{Tncft}). 

 Since $M_{\rm eff}^{\rm induced}\sim -M$ for $r\gg r_0$, 
when $r_h\rightarrow r_0$
from bellow, the black hole absorbs $-M$ units of $D5$ charge and thus the total charge outside the horizon 
$\lim_{r_h\rightarrow r_0} 
M_{\rm eff}^{\rm total}(r_h)\rightarrow 0$. This means
when we consider the case $r_h\ge r_0$, we can neglect three form flux and consider $\widetilde{F}_5$ to be the only
 non-zero flux.
Thus for  $r_h\ge r_0$, we have an exact black hole solution in $AdS_5\times T^{1,1}$ background, with the following warp factors and flux
strength
\bg \label{AdSmetric}
e^{-4A_{\rm AdS}}&=&h_{\rm AdS}=\frac{27\pi\bar{N}_{\rm eff}\alpha'^2}{4r^4}\nonumber\\
e^{2B_{\rm AdS}}&=&1-\frac{r_h^4}{r^4}\nonumber\\
\widetilde{F}_5^{AdS}&=&(1+\ast_{10}) \frac{\partial h^{-1}_{\rm AdS}}{\partial r} dr\wedge dt\wedge dx\wedge dy\wedge dz
\nd   
 where $\bar{N}_{\rm eff}$ is a free parameter related to the five form flux strength,
 \bg
 \bar{N}_{\rm eff}&=&\frac{1}{2 \kappa_{10}^2\mu_3}\int_{T^{1,1}} \widetilde{F}_5^{\rm AdS}
 \nd

 The Wald entropy and the corresponding free energy of  $AdS_5\times T^{1,1}$ is easily computed
 \bg \label{entropyR2}
 s_{\rm R_1+R_2}&=&\frac{\pi r_h^3 \sqrt{27\pi \bar{N}_{\rm eff}}V_8\alpha'}{108\kappa_{10}^2}\nonumber\\
 &=&\frac{27}{32}\pi^2\bar{N}_{\rm eff}^2 V_3 T_{\rm AdS}^3 \nonumber\\
 F_{\rm R_1+R_2}&=&-\frac{27}{128}\pi^2\bar{N}_{\rm eff}^2 V_3 T^4_{\rm AdS}
 \nd 
 In the above, the temperature is defined through (\ref{Tncft}) using the AdS metric (\ref{AdSmetric}):
 \bg\label{TR2}
 T_{\rm AdS}(r_h)=\frac{2r_h}{\pi \sqrt{27\pi \bar{N}_{\rm eff}}\alpha'} 
 \nd

 (\ref{entropyR2}) describes entropy and free energy of the gauge theory for black hole horizons $r_h\ge r_0$ in terms of temperature $T_{\rm AdS}$
  and effective degrees of freedom $\bar{N}_{\rm eff}$. While (\ref{entropyR1},\ref{FreeER1}) describes entropy and free energy of the gauge theory 
  for black hole horizons $r_h<r_0$.  To compare temperatures described by these different size black holes, we must first obtain the exact
  expression for temperature described by  region 1. The
   explicit forms for the metric in region 1 is given by 
  \bg
  h(r)&=&\frac{27\pi N\alpha'^2}{4r^4}\left[1+\frac{g_sM^2}{N}{\rm log}\left(\frac{r}{r_*}\right)+\frac{g_sM^2}{N}c_l 
  \left(\frac{\tilde{r}_h}{r}\right)^{4l}\right]\nonumber\\
  g(r)&=&1-\frac{\tilde{r}_h^4}{r^4}+\frac{g_sM^2}{N}d_l 
  \left(\frac{\tilde{r}_h}{r}\right)^{4l}
  \nd
  where $c_l,d_l$ are constants independent of $M,N$. The horizon radius $r_h$ is such that $g(r_h)=0$, which gives that $r_h$ is the solution of the following equation 
  \bg
   \frac{\tilde{r}_h^4}{r_h^4}&=&1+ \frac{g_sM^2}{N}d_l 
  \left(\frac{\tilde{r}_h}{r_h}\right)^{4l}
  \nd
  Putting everything together, we get that the temperature described exclusively by region 1 without the inclusion of region 2 is
  \bg \label{TR2}
  T(r_h)&=&\frac{2r_h\left(1+\frac{g_sM^2}{N}d_l (1+l) 
  \left(\frac{\tilde{r}_h}{r_h}\right)^{4l}\right)}{\pi \sqrt{27\pi N}\alpha'\left[1+\frac{g_sM^2}{N}{\rm log}\left(\frac{r_h}{r_*}\right)+\frac{g_sM^2}{N}c_l 
  \left(\frac{\tilde{r}_h}{r_h}\right)^{4l}\right]^{1/2}}
  \nd
  
  As $r_h\rightarrow r_0^+$, black holes in
  $AdS_5\times T^{1,1}$ geometry should describe the same temperature as the black holes in region 1, with $r_h\rightarrow r_0^-$. This means
  we must have 
  \bg\label{Tcond}
  T_{\rm AdS}(r_h \rightarrow r_0^+)=T(r_h \rightarrow r_0^-)\equiv T_0
  \nd
  The above condition can be used to relate $\bar{N}_{\rm eff}$ with $N$ to give
  \bg \label{matchingcond}
  \bar{N}_{\rm eff}&=& \frac{N\left[1+\frac{g_sM^2}{N}{\rm log}\left(\frac{r_0}{r_*}\right)+\frac{g_sM^2}{N}c_l 
  \left(\frac{\tilde{r}_h}{r_0}\right)^{4l}\right]}{1+\frac{g_sM^2}{N}d_l (1+l) 
  \left(\frac{\tilde{r}_h}{r_0}\right)^{4l}}\nonumber\\
  &\equiv&N\left[1+\frac{2e_0g_sM^2}{N}+\frac{2e_1g_sM^2}{N}{\rm log}\left(\frac{r_0}{r_*}\right)\right]
  \nd
   where we have introduced constants $e_0,e_1$ independent of $M,N$ and determined by above equation. Expanding only up to linear order in
   ${\cal O}(\epsilon)$ and using (\ref{matchingcond}) in (\ref{entropyR1},\ref{entropyR2}), we readily get the entropy difference 
  \bg \label{dels}
  \triangle s&=&s_{\rm R_1+R_2}(r_h\rightarrow r_0^+)-s_{\rm R_1}(r_h\rightarrow r_0^-)\nonumber\\
  &=&\frac{\pi r_0^3 \sqrt{27\pi N}V_8\alpha'}{108\kappa_{10}^2}\left(\frac{g_sM^2}{N}\left[e_0-a_0\right]+\frac{g_sM^2}{N}
  \left[e_1-a_1\right]{\rm log}\left(\frac{r_0}{r_*}\right)\right)
  \nd
  
  As entropy always increases with temperature, we must have $\triangle s\ge 0$, which will automatically lead to 
  \bg
  \triangle F=F_{\rm R_1+R_2}(r_h\rightarrow r_0^+)-F_{\rm R_1}(r_h\rightarrow r_0^-)<0
  \nd
  
  Finally we can compute the internal energy and pressure and evaluate the conformal anomaly. At low energies for the case 
  $d> \alpha^1-1/2$ and ignoring ${\cal I}$,
  we can have a regime $ T_c\le T \le T_0$. Then
   the internal energy and
  pressure is given by black hole in region 1 and we get 
  
  \bg
  e_{\rm R_1}&=&\frac{1}{V_3}\left(F_{\rm R_1}+Ts_{\rm R_1}\right)\nonumber\\
  &=&\frac{27\pi^6 T^4 N^2V_5\alpha'^4}{128\kappa_{10}^2}\left(3+\frac{3b_0 g_sM^2}{N}+\frac{b_1 g_sM^2}{4N}+
  \frac{3b_1 g_sM^2}{N}{\rm log}\left(T\sqrt{N\alpha'}\right)\right)\nonumber\\
  p_{\rm R_1}&=&-\frac{\partial F_{\rm R_1}}{\partial V_3}
 \nd
 This readily gives the conformal anomaly 
 \bg
 \triangle_{\rm R_1}\equiv \frac{e_{\rm R_1}-3p_{\rm R_1}}{T^4}=\frac{27\pi^6  N^2\alpha'^4V_5b_1 g_sM^2}{512\kappa_{10}^2N}
 \nd
 On the other hand, with $T>T_0$, the gauge theory is described by black hole in $AdS_5\times T^{1,1}$, where 
 \bg
 \triangle_{\rm R_1+R_2}=\triangle_{\rm AdS}=0
 \nd
 While for $T<T_c$, for the confined phase, we can take both internal energy and pressure to be zero to obtain $\triangle_{\rm confine}=0$. 
  
\begin{figure}[htb]\label{gauge}
       \begin{center}
\includegraphics[height=6cm]{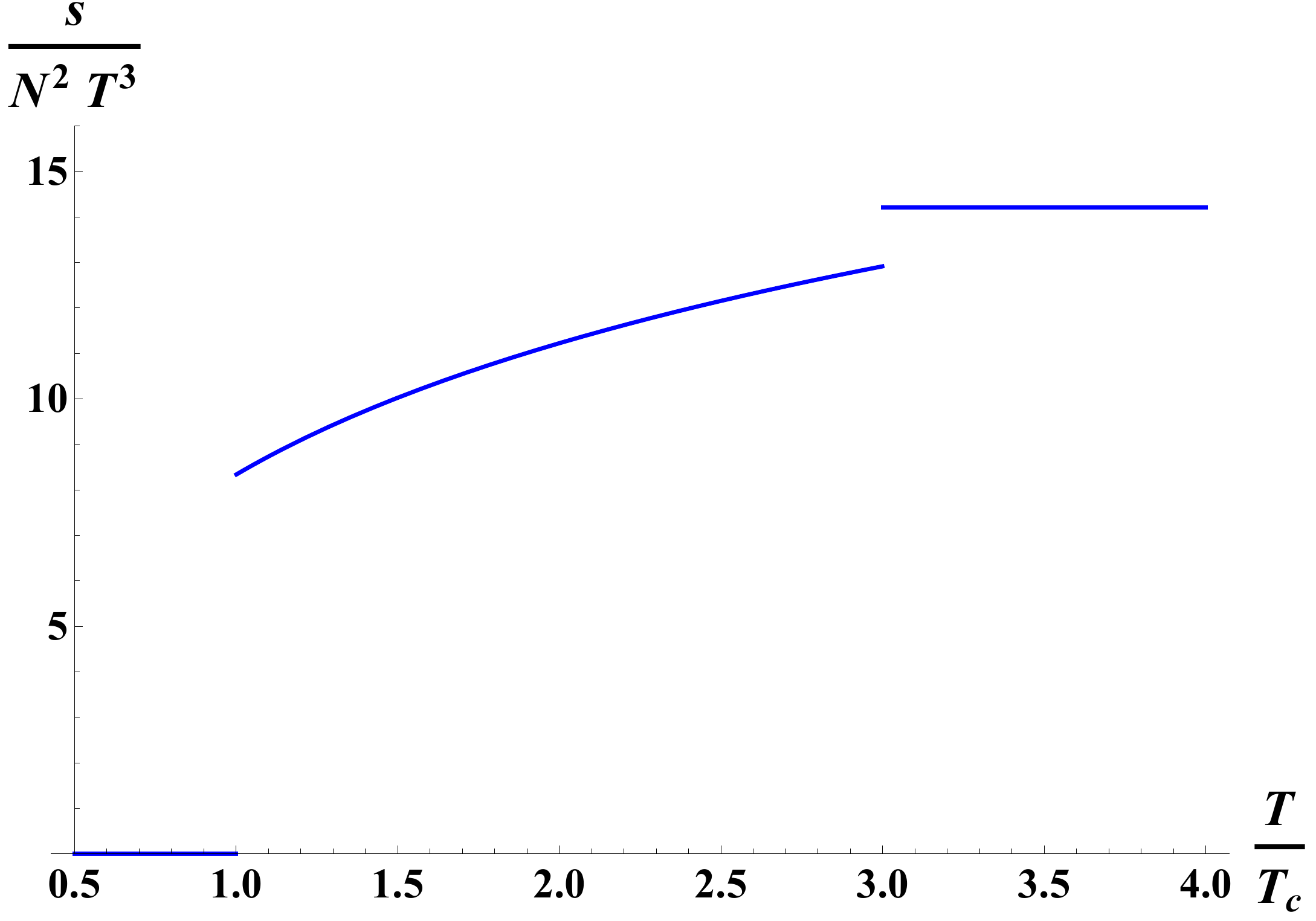}
        \caption{$\frac{s}{N^2T^3}$  as a function of $\frac{T}{T_c}$ with boundary condition $d-\alpha^1=\frac{4{\rm
	log}(3)}{18}-\frac{1}{2}$.}
       \end{center}
        \end{figure}
	\begin{figure}[htb]\label{gauge}
       \begin{center}
\includegraphics[height=6cm]{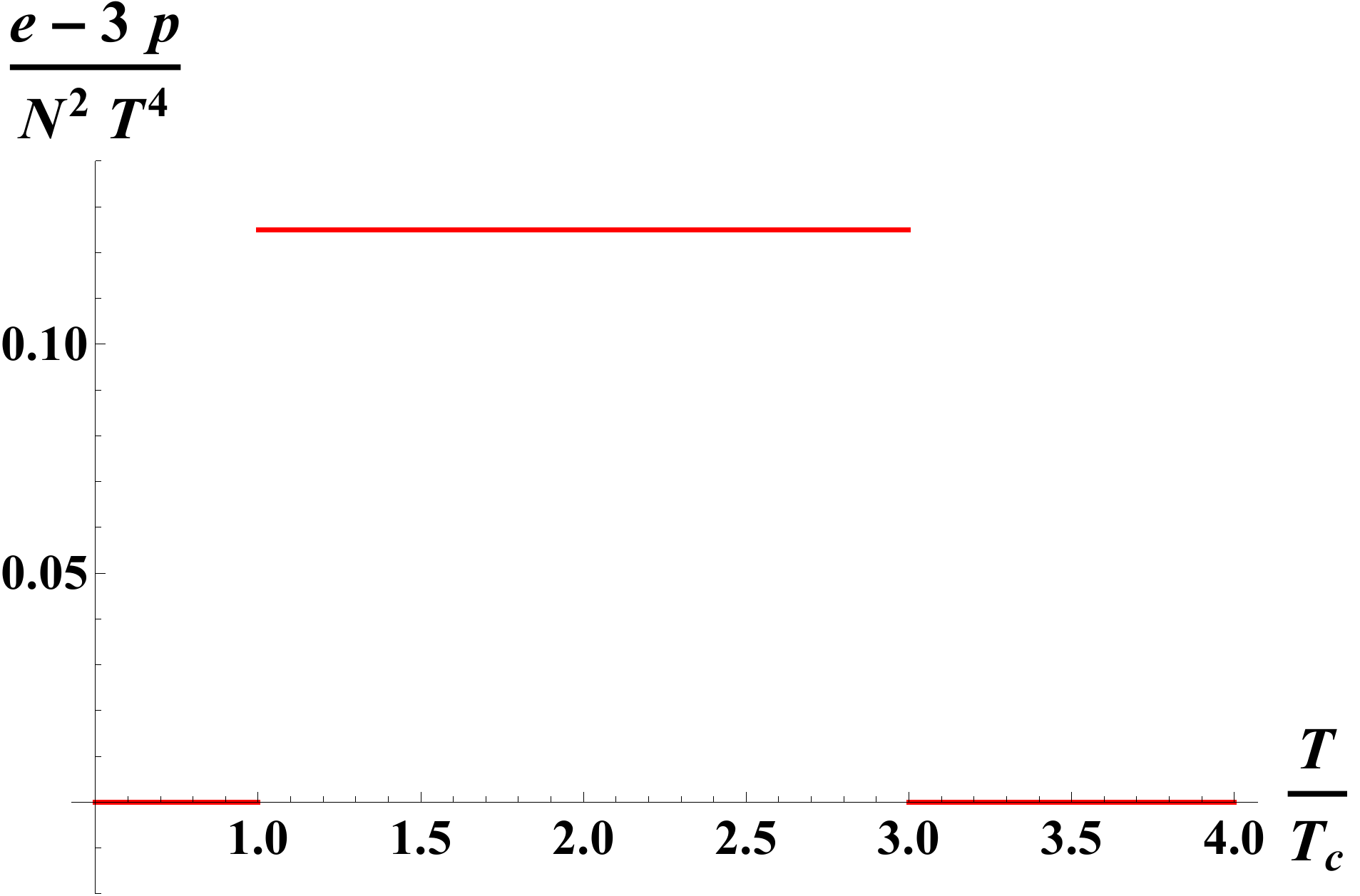}
        \caption{$\triangle$  as a function of $\frac{T}{T_c}$ with boundary condition $d-\alpha^1=\frac{4{\rm
	log}(3)}{18}-\frac{1}{2}$.}
       \end{center}
        \end{figure}
 
 We can summarize the scenario as follows:

\noindent $\bullet$ $d> \alpha^1-1/2$: The low energy regime of the gauge theory is given by region 1 and we get a first order Hawking-Page 
like transition with critical temperature $T_c$ given by (\ref{Tcc}). In deriving $T_c$, we have ignored ${\cal I}$, which is only
valid if $r_h^c\ll r_0$. On the other hand, if $r_h^c$ is too small, our perturbative analysis is invalid since there is no
small perturbative parameter. Thus (\ref{Tcc}) is only an approximation which gets more and more accurate if $r_h^c$ is 
in narrow  regime were
${\cal I}$ is negligible and the perturbative parameter (\ref{ppara}) exists. 

For $T<T_c$ we have confinement while $T>T_c$ describes a deconfined phase. 
 Region 1 describes all temperatures up to $T_0$ and the
entropy of the gauge theory is given by (\ref{entropyR1}). 
To incorporate the high temperature modes $T>T_0$  of the gauge theory, we  add a
UV cap to the geometry by considering back reactions of localized world volume fluxes sourced by on D7 branes. Then the entropy is given by  
(\ref{entropyR2}). 

By using relation between horizon and temperature, we can obtain entropy as a function of temperatures for all $T$. The
result is plotted in Fig 6. Whatever the boundary condition, we must satisfy the thermodynamic condition (\ref{dels}) and 
 Fig 6 is the scenario where  $\triangle s>0$, both across $T_c$ and $T_0$. The jump at $T_c$ corresponds to the first order
 Hawking-Page transition while the jump at $3T_c=T_0$ is consistent with the scenario that $D7$ brane has 
 fallen into the black hole, resulting in increased entropy. This discontinuity at $T_0$ can also be understood in the
 following way: For $T<T_0$, we are using the gravity action $S_{\rm R_1}$ of region 1, neglecting the geometry of region 2
 with localized sources. Region 1 with a black hole is holographic dual to a gauge theory at thermal equilibrium.
 Only considering the action $S_{\rm R_1}$ is equivalent to the scenario that this
 gauge theory does not interact with localized sources of region 2. When $T>T_0$, gauge theory dual to region 1 comes in
 thermal equilibrium with the localized sources of region 2 and the to total system is described by an 
 $AdS_5\times T^{1,1}$ black hole. We expect the entropy of the combined system to be larger than the entropy gauge theory
 dual to region 1 and thus the discontinuity at $T_0$ is reasonable. In Fig 7, we have plotted the anomaly,which is just a box of width $3T_c$.  

\noindent $\bullet$ $d\le \alpha^1-1/2$: Again, low energy regime of the gauge theory is given by region 1. However now, for all 
$T<T_0$ only
the geometry $X^1$ describes non-zero temperature and we have the confined phase. For $T>T_0$, we add region 2 and  only the black 
hole geometry describes non-zero
temperature with entropy given by (\ref{entropyR2}). That is we have thermal CFT for $T>T_0$ and a confined theory for $T<T_0$. Thus for the
condition $d\le \alpha^1-1/2$, we can treat $T_0$ as the critical temperature of deconfinement. Note that we are ignoring 
${\cal
I}$ and using our perturbation in ${\cal O}(\epsilon)$, both of which is only valid if $r_h$ is a narrow regime.

The conformal anomaly of Fig 7 looks very different from what is expected in QCD. The discrepancy can be attributed to the
break down of our perturbative analysis as follows:

\noindent $\bullet$ The width of $3T_c$ is obtained by choosing  $d=\alpha^1+\frac{4{\rm
	log}(3)}{18}-\frac{1}{2}$. This value of $d$ could be too large such that $r_h^c$ is too small and our perturbative
	analysis in ${\cal
	O}\left(\epsilon \right)$ breaks down. Thus it is possible that for an anomaly of 
	width $3T_c$, the computation of anomaly need to be modified.

\noindent $\bullet$ It is possible that $r_h^c$ is large such that we cannot ignore ${\cal I}$. The derivation of $T_c$ ignores
${\cal I}$ entirely and for $r_h\lesssim r_0$, ${\cal I}$ can be large enough to invalidate the computation of $T_c$. Although large
horizon makes our perturbation in ${\cal O}\left(\epsilon\right)$ exact, it also makes ${\cal I}$ large and thus
computation of $T_c$ less and less exact. In other words, if we accounted for ${\cal I}$, then the computation of $T_c$ along with
the free energy of the black hole in region 1 could get significant modifications. These modifications could drastically alter the
anomaly and bring it closer to what is seen in lattice QCD. 

In light of the above discussions, it would be ideal to consider small black holes, for which ${\cal I}$ can be neglected. However
the fluxes in section 2 were computed using a perturbative series in ${\cal O}(\epsilon)$, where we only kept up to linear order
terms in the action. For small black holes, fluxes cannot be written as such a series since there is no small parameter like
(\ref{ppara}). In other words, if we were to write the fluxes as a series, the relevant parameter is then  
$\tilde{\epsilon}\equiv
\frac{g_sM^2}{N(\rho_c)}$. For small black holes, $\rho_c$ will be small and 
$\tilde{\epsilon}$ will be large. Then higher order terms in  $\tilde{\epsilon}$ are equally or more important 
than the linear order term 
and thus
we cannot exactly compute the on-shell gravity action.  
Consequently we cannot obtain the thermodynamics near critical temperature. On the other hand, small black holes are dual to QCD like gauge theories near critical temperature. Thus in the following section
we consider small black holes in deformed warped cone which in fact give rise to QCD like features.  

\subsection{Connection to QCD}
As already discussed in the previous sections, the perturbative analysis used to derive the black hole solutions and the
resulting thermodynamic state functions of the gauge theory breaks down when the horizon $r_h$ is small. However, small
horizon could give rise to temperatures that are above or near the deconfinement temperature of QCD. This is because large
black holes describe  ultra violate modes of dual gauge theories with gauge group $SU(N+M)\times SU(N)$, which is nothing like
QCD. Only at low energy the
gauge group cascades down to $SU(M)$, which can resemble QCD.  Thus to determine the
thermodynamics of the strongly coupled gauge theory that behaves like QCD, we must obtain black hole solutions with 
$r_h$ small. 

On the other hand, when $r>r_h$ is small, the internal metric (\ref{inmate}) {\it does not} take the simple form
(\ref{inmate1}) of a regular cone and  we need to obtain black holes in warped {\it deformed cone} where the radial coordinate
is $\rho$ and horizon is $\rho_h$. When there is no horizon and we restrict to region 1, where there is no localized source, 
the internal metric is given
by $\tilde{g}_{mn}^0$. When horizon $\rho_h\neq 0$, the internal metric is $\tilde{g}_{mn}=\tilde{g}_{mn}^0+\tilde{g}_{mn}^1$
with $\tilde{g}_{mn}^1\neq 0$. For very large horizons, $\tilde{g}_{mn}^1$ is small and given by (\ref{gmn}), but for small
horizons the higher order terms are equally or more important than the leading term in the expansion and we do not have exact
expressions. Without knowing the form of $\tilde{g}_{mn}^1$, we cannot find the horizon which is obtained by solving
(\ref{BHfactorA}). 

However, we can still obtain the form of the function $B(x^m)$ by first noting that key quantity that enters (\ref{BHfactorA})
and crucially depending on  $\tilde{g}_{mn}$ is 

\bg
{\cal H}_n\equiv \tilde{g}^{pq}\partial_n \tilde{g}_{pq}
\nd
where $p,q=4,..,9$ runs over the cone directions.
If $n=\rho$ and we use $\tilde{g}_{pq}=\tilde{g}_{pq}^0$, then one readily gets that ${\cal H}_\rho={\cal H}_\rho(\rho)$ is
{\it only} a function of $\rho$. Then, we can solve (\ref{BHfactorA}) to obtain that $B(\rho)$ is only a function of $\rho$.     
This leads to a horizon that is an eight dimensional surface described by $\rho=\rho_h$ where $e^{B(\rho_h)}=0$. In region 1,
there are no localized sources and no source for asymmetry.  Thus we expect the horizon to be a surface given by 
$\rho=\rho_h$ even when $\tilde{g}_{pq}^1$ corrections are considered and
$\tilde{g}_{pq}=\tilde{g}_{pq}^0+\tilde{g}_{pq}^1$. Thus including the metric corrections  $\tilde{g}_{pq}^1$, we expect

\bg
\tilde{g}^{mn} \tilde{g}^{pq}\partial_n \tilde{g}_{pq}\partial_m B &=&\tilde{g}^{\rho\rho}
\tilde{g}^{pq}\partial_\rho \tilde{g}_{pq}\partial_\rho B\nonumber\\
&=&\left(\tilde{g}^{\rho\rho,0}
\tilde{g}^{pq,0}\partial_\rho \tilde{g}_{pq}^0+{\cal Q}(\rho)\right)\partial_\rho B
\nd   
 where ${\cal Q}(\rho)$ arises due to corrections $\tilde{g}_{pq}^1$. If we further assume
 $\tilde{g}_{\rho\rho}=\tilde{g}^0_{\rho\rho}$ .i.e the perturbations $\tilde{g}_{pq}^1$ are only in the compact direction,
 (\ref{BHfactorA}) drastically simplifies to give
 
 \bg \label{BHeqrho}
 &&\left[32 \rho {\rm cosh}(\rho) + 4 \rho {\rm cosh}(3\rho) - 5 {\rm sinh}(\rho) - 12 {\rm sinh}(3 \rho) + 
    {\rm sinh}(5 \rho)+\tilde{{\cal Q}}\right] g'(\rho) \nonumber\\
    &&+ 
 2 \left[-2 \rho + {\rm sinh}(2 \rho)\right] {\rm sinh}(3 \rho)g''(\rho)=0 
 \nd 
 where prime denotes a derivative with respect to $\rho$ and $\tilde{{\cal Q}}\neq 0$ only when $\tilde{g}_{pq}^1\neq 0$. We
 need to solve the above equation along with (\ref{warp_eq_1}), and (\ref{GKP_BH}) to obtain the scalar functions $g(\rho),
 h(\rho)$ and $\gamma(\rho)$.   
 When  we are in large $\rho$
 region such that $\tilde{g}_{mn}^0$ is given by (\ref{inmate1}), the above equation has a simple solution in $r$ coordinate
as written in (\ref{alpha}). For small $\rho$, there are corrections to $g(r)$ and we can write 

\bg \label{grho}
g(\rho)=1-{\rm exp}\left(\frac{4\rho_h^0}{3}-\frac{4\rho}{3}\right)+G(\rho)
\nd         
where $\rho_h^0$ is a constant related to $\tilde{r}_h$.
Plugging in the above form (\ref{grho}) in (\ref{BHeqrho}), we can obtain a second order linear differential equation for
$G(\rho)$, which can be exactly solved if $\tilde{{\cal Q}}(\rho)$ was known. We can numerically solve (\ref{BHeqrho}) by
plugging in a  Taylor series for $\tilde{{\cal Q}}$
\bg
\tilde{{\cal Q}}=q_i \rho^i
\nd   
where $q_i$ are constants. In fig 8 we plotted $g(\rho)$ obtained by solving (\ref{BHeqrho}) with $\tilde{{\cal Q}}=0$ and the
boundary condition $G(100)=0,G'(10)=\frac{5}{10000}$ with the choice $\rho_h^0=1$. 
The solution is dependent on these boundary conditions and the
choice for $\tilde{{\cal Q}}$ and is only presented to demonstrate the qualitative feature of $g(\rho)$. Of course the Einstein
equations in (\ref{Ricci_Min}) will determine $\tilde{g}_{mn}^1$ and consequently $\tilde{{\cal Q}}$. 
But since we do not have a
solution to the set of flux equations and Einstein equations for small $\rho$, we only present a numerical solution to
(\ref{BHeqrho}) ignoring $\tilde{{\cal Q}}$ to understand the form of the function $g(\rho)$. Solution to (\ref{Ricci_Min})
will give $q_i$ which are not necessarily zero, and thus the numerical solution presented in the plot will be altered.   
\begin{figure}[htb]\label{gauge}
       \begin{center}
\includegraphics[height=6cm]{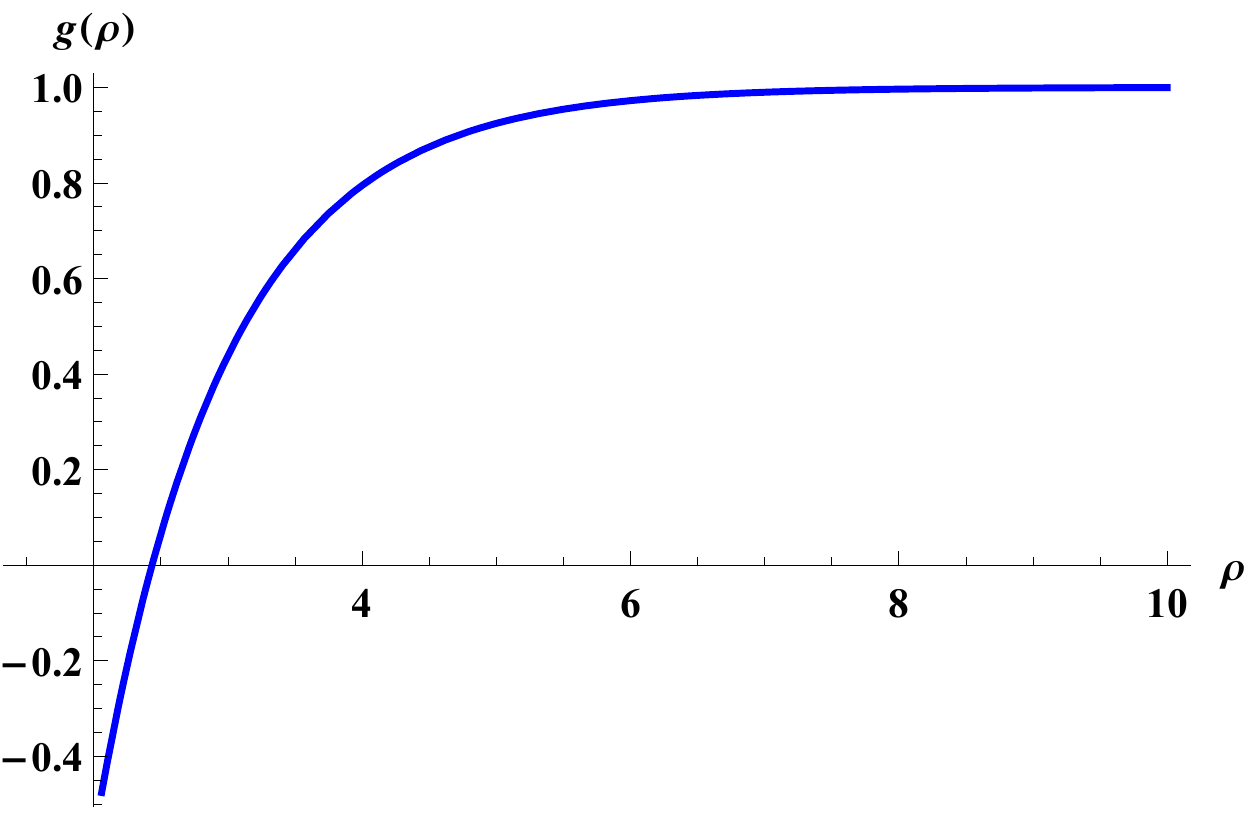}
        \caption{$g(\rho)$ as a function of $\rho$. The zero of the function gives the horizon $\rho_h$.}
       \end{center}
        \end{figure}
From our numerical solution, we find that the horizon is at 

\bg
\rho=\rho_h\simeq 2.5 \rho_h^0
\nd

Now, to obtain temperature, we need to find the warp factor for small $\rho$ in the presence of a black hole. We need to solve (\ref{warp_eq_1})
and (\ref{GKP_BH}) with $g(\rho)$ given by (\ref{grho}). When $\rho_h=0$, $h(\rho)$ is given by the Klebanov-Strassler solution
\bg
h_{\rm KS}(\rho)=\alpha_0 \frac{2^{2/3}}{4}\int_{\rho}^\infty d\zeta \frac{\zeta {\rm coth}(\zeta)-1}{{\rm
sinh}^2(\zeta)}\left({\rm sinh}(2\zeta)-2\zeta\right)^{1/3}
\nd
In the presence of the black hole, this solution is altered and we expect a regular solution
\bg \label{hrho}
h(\rho)=h_{\rm KS}(\rho)+\tilde{c}_i\rho^i
\nd
where $\tilde{c}_i$ are constants such that $h(\rho_h)\neq 0$. 

With the form of $g(\rho)=e^{2B}, h(\rho)=e^{-2A}$ known, we can find the temperature 
\bg\label{Trho}
T(\rho_h)=\frac{g'(\rho_h)}{4\pi \sqrt{\tilde{g}_{\rho\rho}(\rho_h)h(\rho_h)}}
\nd
In Fig 9, we have plotted the points $T(\rho_h)$ for various values of  horizon $\rho_h$. 
In deriving the points shown in red dots,
we numerically solved (\ref{BHeqrho}) for various choices of $\rho_h^0$ i.e $\rho_h^0=1,2,2.2,...,6$, every time 
solving the
differential equation with the condition $G(100)=0,G'(10)=0$. Each solution give a  value for the horizon $\rho_h$ and then
for each $\rho_h$, we evaluated (\ref{Trho}) using $h(\rho)\sim h_{\rm KS}(\rho)\sim \frac{3}{4}2^{1/3}(g_sM)^2 \rho
e^{-4\rho/3}$, which is again an approximation. This approximation gets better for larger $\rho$ and since $\rho>\rho_h>2$,
we expect the approximation to be a reasonable one. Also note that we are ignoring $\tilde{c}_i$ since we do not know their
exact values. Of course solving the Einstein equations will determine the constants $\tilde{c}_i$ and the solution for
$T(\rho_h)$ will depend on $\tilde{c}_i$. However, we expect the form of  $T(\rho_h)$ to remain unchanged, that is it
should  behave as a Taylor series 

\bg
T(\rho_h)=t_i\rho_h^i
\nd   
where $t_i$ are constants. 
For our plot in Fig 9, we were able to fit all the points with $T=0.234-0.05\rho_h+0.126\rho_h^2$ and the fit is shown in
the blue curve. Of course, when $\tilde{c}_i$ are
included, the coefficients will change. In generating this plot, we have set the constants such that 
\begin{figure}[htb]\label{gauge}
       \begin{center}
\includegraphics[height=8
cm]{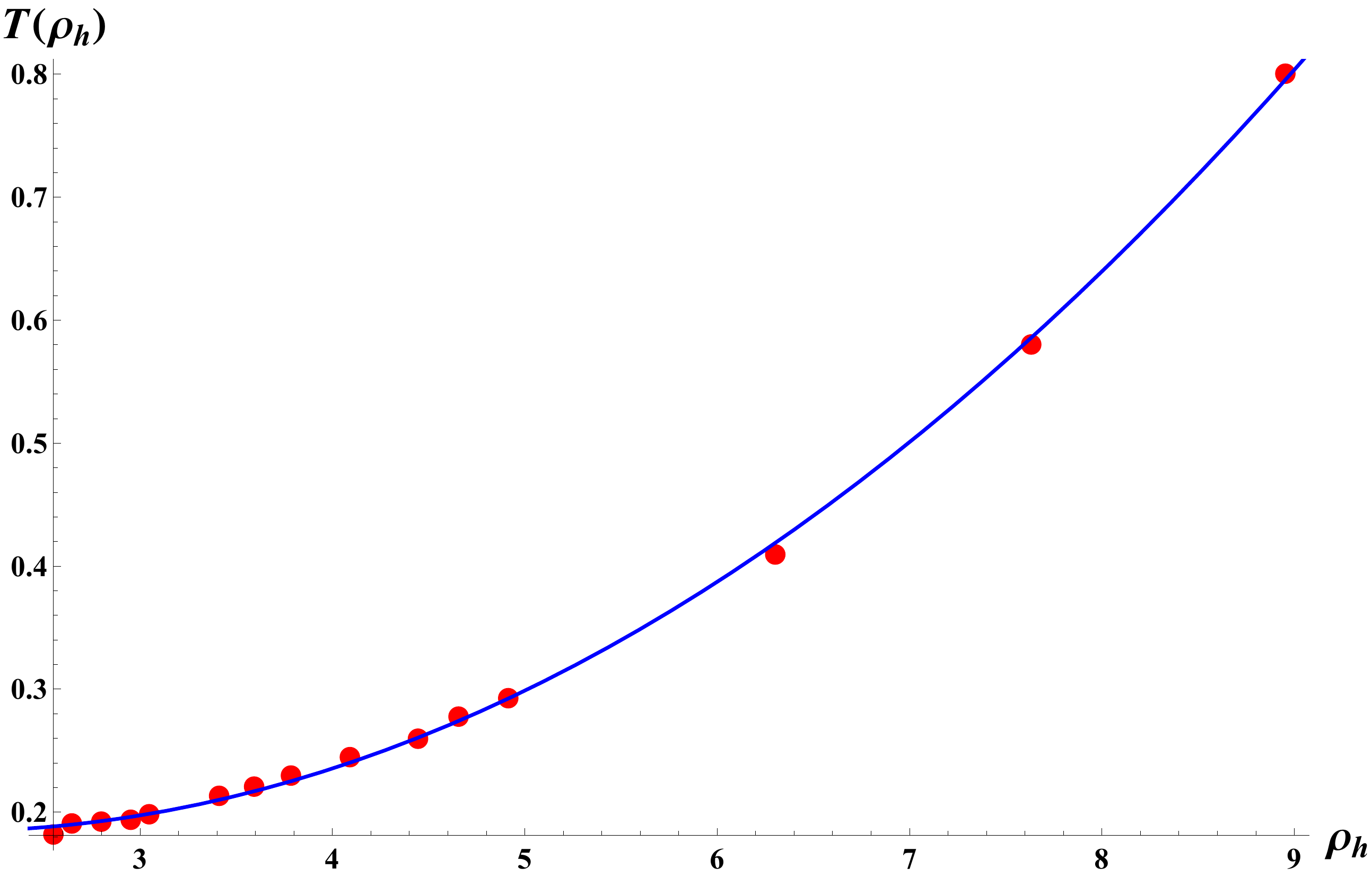}
        \caption{$T(\rho_h)$ as a function of $\rho_h$. Solid line is the polynomial fit while the points are generated using
	the numerical solution.}
       \end{center}
        \end{figure}
\bg
g_s^2M^2=\frac{16}{{\cal A}^{4/3}2^{1/3}}
\nd 
which can lead to large $g_sM$ for a relatively small value of ${\cal A}$. Also, to simplify and avoid keeping track of
$\alpha'$, we set $\alpha'=1$ in obtaining the plots and in what follows.

Using Wald's formula, we can find the entropy of the black hole, 
\bg \label{entropyrho}
s\sim \sqrt{h(\rho_h)}{\rm sinh}(\rho_h)\left({\rm sinh}(2\rho_h)-2\rho_h\right)^{1/3}
\nd
\begin{figure}[htb]\label{gauge}
       \begin{center}
\includegraphics[height=6cm]{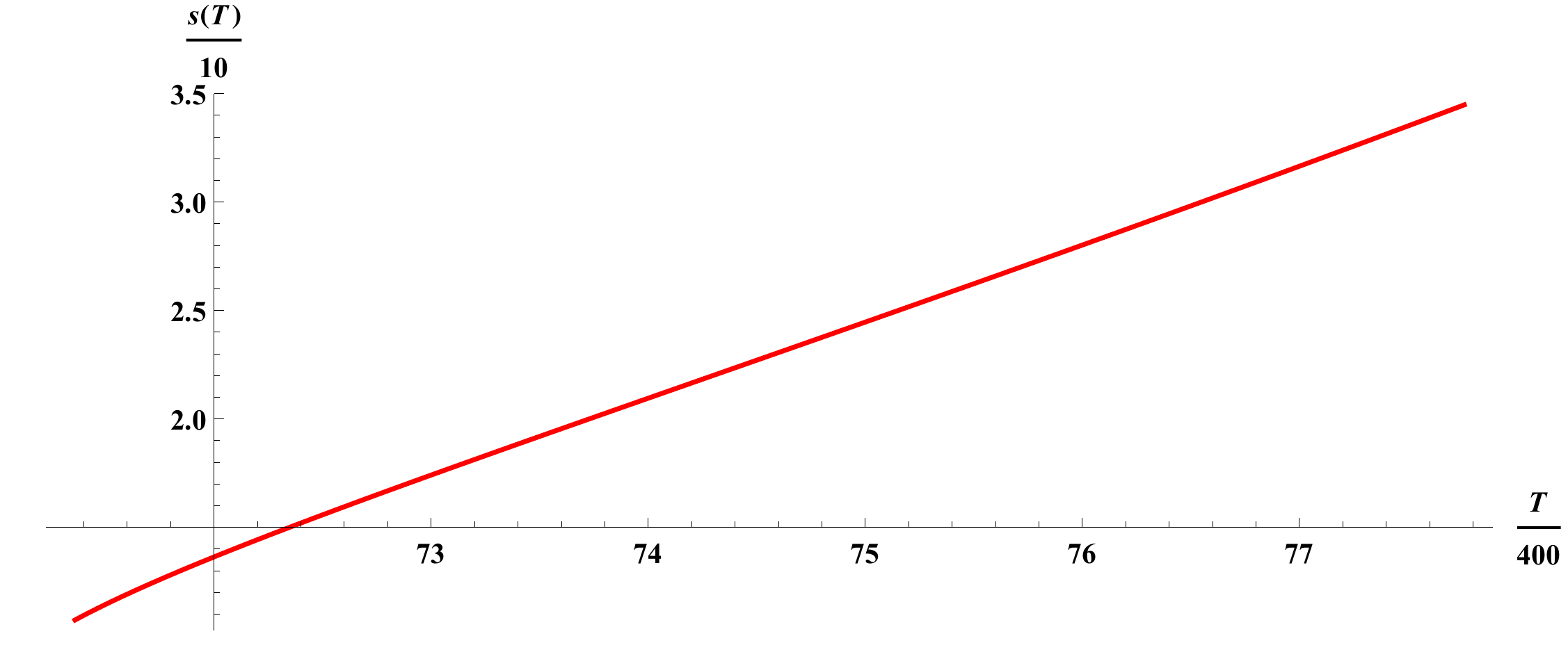}
        \caption{$s(T)$ as a function of $T$, for small $T$. }
       \end{center}
        \end{figure}
	\begin{figure}[htb]\label{gauge}
       \begin{center}
\includegraphics[height=7cm]{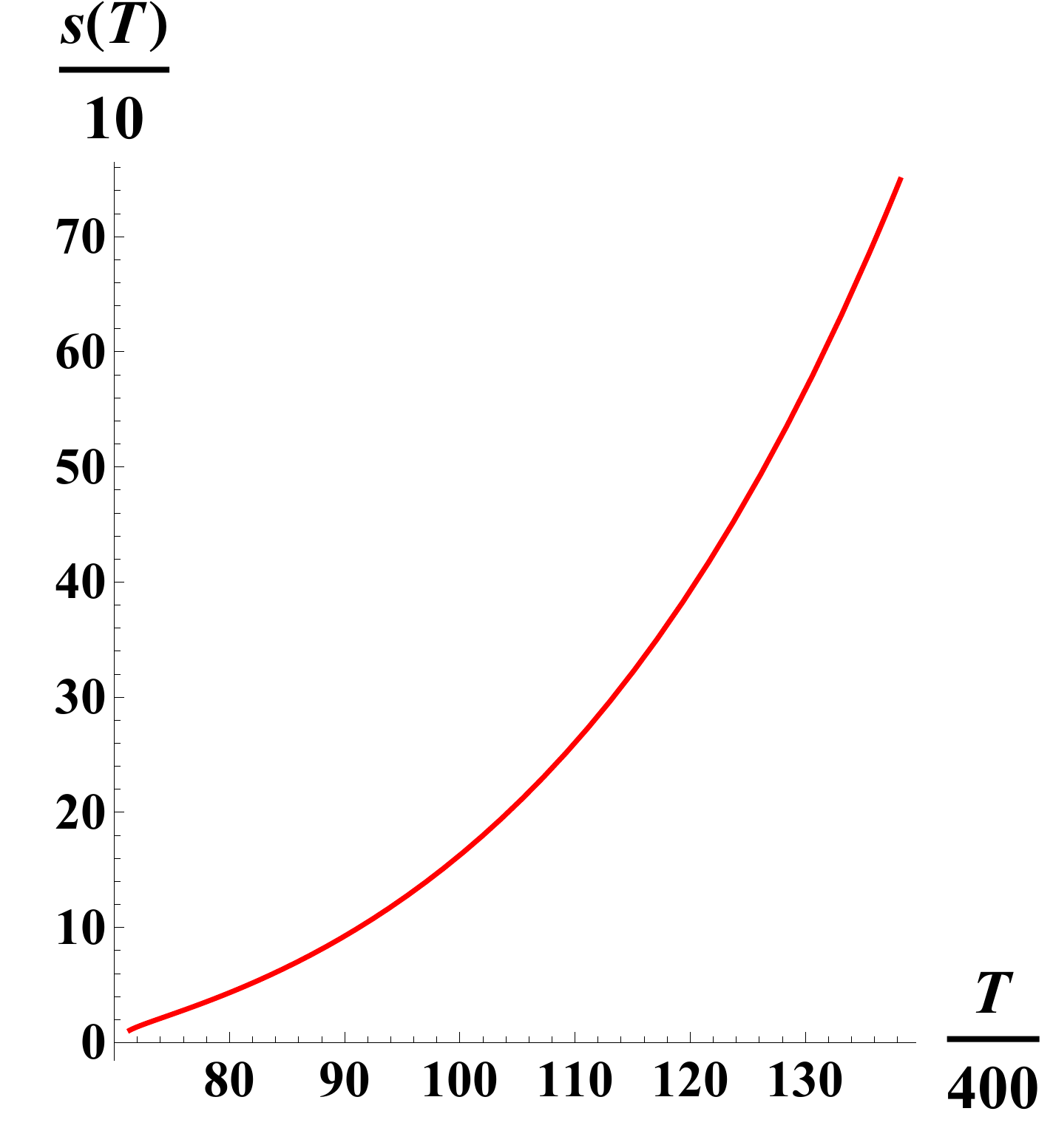}
        \caption{$s(T)$ as a function of $T$, for large $T$.}
       \end{center}
        \end{figure}
Both $s$ and $T$ are functions of horizon $\rho_h$ and thus we can plot $s$ as a function of $T$. 
This is done in  Fig 10 and 11 where temperature is obtained from the fit in Fig 9 and entropy is obtained using the
scaling in (\ref{entropyrho}). For
small $T$, we see that $s\sim T$ while for larger $T$, $s\sim T^2$. These scalings arise by considering black holes in deformed
cone, that is for small black holes. On the other hand for large black holes i.e. $r_h$ large, the
entropy and free energy scales as (\ref{sfT}), (\ref{FreeER1}) and (\ref{entropyR2}). Combining all these results, we find
 that the black hole has the
following scaling of free energy with temeprature

\bg \label{F-R1}
F&\sim& -T^2  ~~~~{\rm for \; small \; T_0>T>T_c}\nonumber\\
&\sim&-T^3~~~~
{\rm for \; intermediate \; T_0>T>T_c}\nonumber\\
&\sim&-T^4\left(1+\frac{b_0 g_sM^2}{N}-\frac{b_1 g_sM^2}{4N}
+\frac{b_1 g_sM^2}{N}{\rm log}\left(T\sqrt{N\alpha'}\right)\right) {\rm for \; large \; T\lesssim T_0}\nonumber\\
&\sim& -T^4~~~~{\rm for \; large \; T>T_0}
\nd

Finally using these scalings, we get the conformal anomaly for the dual gauge theory,
\bg\label{Anomaly-R1}
\triangle&\sim& \frac{1}{T^2} ~~~{\rm for \; small \; T_0>T>T_c}\nonumber\\
&\sim& \frac{1}{T}~~~~~{\rm for \; intermediate \; T_0>T>T_c}\nonumber\\
&\sim& \frac{27\pi^6  N^2\alpha'^4V_5b_1 g_sM^2}{512\kappa_{10}^2N}~~{\rm for \; large \; T\lesssim T_0}\nonumber\\
&\sim& 0~~~~~~~{\rm for \; large \; T>T_0}
\nd

Since we do not
have exact numerical solutions for small black holes in region 1 (we only have exact actions for large black holes in region 1),
we cannot compute exact on-shell action. Thus we cannot obtain
the critical temperature $T_c$ above which black holes in deformed cone are preferred over vacuum. However, we have obtained the
scalings of free energy, entropy and conformal anomaly for the black holes which can be done without computing $T_c$. If $T_c\ll
T_0$ is
small, the deconfined phase of the gauge theory has the scalings given by (\ref{F-R1}, \ref{Anomaly-R1}), which is qualitatively similar to the
lattice QCD simulations. On the other hand, the value of $T_c$ depends on  values of the metric near the horizon and boundary
$r=r_b$ and these boundary conditions can be altered. Thus by picking a certain class of boundary conditions, we expect to obtain
$T_c\ll T_0$ and thus the scalings can describe the deconfined phase of a gauge theory that arises from the brane
configuration of Fig 3. This way, we can obtain a black hole description for a thermal gauge theory that confines at IR, becomes
conformal at the UV and behaves similar to QCD near $T_c$.

 \section{Conclusion}
 In this paper we demonstrated how the UV divergence of Klebanov-Strassler (KS) model can be eliminated by considering world volume
 fluxes on $D7$ branes.  The back reaction of the world volume fluxes modifies the KS fluxes $F_3,H_3$ for large $r$ and the resultant fluxes are
 explicitly given by (\ref{G3solu1}). These fluxes are evaluated in the presence of a black hole and by taking the limit
 $\tilde{r}_h\rightarrow 0$, we can obtain the the fluxes in vacuum. 
 
 Using these fluxes, we then showed that the effective $D5$ charge in the dual gauge theory vanishes in the far UV. The metric 
 and fluxes are
 similar to the KS model in small `r' region and thus the gauge theory confines in far IR. Hence we end up with a gauge theory that
 is UV conformal and IR confining- similar to QCD.    The RG flow of the gauge theory can be
 directly obtained by using (\ref{RGflow}) with fluxes and dilation field given by (\ref{B2}), (\ref{Imtau}). A more detailed
 analysis of RG flow will be presented in our upcoming work \cite{Meson-spec}.
 
 The presence of the localized $D7$ branes along with world volume fluxes introduced an additional scale $r_0$ in the theory. The
 fluxes and dilaton field we obtained are identical to the KS model (up to linear order ${\cal O}(\epsilon)$) in region 1 i.e. $r<r_0$ while they are modified in region 2
 i.e. $r>r_0$. Thus $r_0$ can be thought of as the radial scale up to which KS solution can be used to study the gauge theory. This
 was done in section 3.1 and 3.3 
 by considering the gravity action $S_{\rm R_1}$ for region 1 and neglecting the localized sources. 
 Although in region 1 we neglected
 the localized sources, which was crucial in analyzing Hawking-Page transition, the scale $r_0$ arising from the sources 
 explicitly entered our analysis. In particular the critical
 temperature $T_c$ given in (\ref{Tcc}) explicitly depends on $r_0$ and thus the localized sources implicitly effect the 
 thermodynamics of the  gauge theory. By considering the black hole entropy in region 1, we were able to
 compute thermodynamic state function of the gauge theory. Since we obtain identical RG flow irrespective of which gravitational
 action we use to describe region 1 ($S_{\rm R_1}$ or $S_{\rm SUGRA}+S_{D7}$), we expect that Wald entropy obtained from 
 $S_{R_1}$ will be similar to the one obtained from $S_{\rm SUGRA}+S_{D7}$. This is reasonable since the entropy depends on the near
 horizon behavior of the metric and both actions  $S_{\rm R_1}$ and $S_{\rm SUGRA}+S_{D7}$ can result in identical horizon values.
          
 In section 3.2 we studied the thermodynamics of the gauge theory when region 2 and localized sources are included. When $D7$
 branes is outside the horizon and we consider back reaction, we argued that the dual gauge theory is not at thermal equilibrium
 and there is no unique temperature. In this scenario, there are no Hawking-Page transitions between geometries. On the oher hand,
 when we consider larger horizons, the $D7$ brane along with world volume fluxes fall into the black hole. This black hole which
 absorbed the $D7$ branes corresponds to a gauge theory at thermal equilibrium and the temperature is given by the Hawking
 temperature of the black hole. We argue for large horizons, we will end up with Schwarzchild black holes since the total $D5$
 charge will be neutralized. Thus at large temperatures, we will obtain a thermal CFT with a dual description in terms of
 Schwarzchild black holes in $AdS_5\times T^{1,1}$. 
 
 Finally in section 3.3, we try to make connections to QCD by considering small black holes in deformed cone geometry. Using the
 form of the warp factors near the tip of the deformed cone,  we worked out a 
 numerical solution for the black hole. Then computing the entropy and temperature of such a black hole, we were able to obtain
 the scaling of conformal anomaly. This scaling is qualitatively similar to QCD near critical temperature. For an exact analysis,
 we need to numerically obtain  black hole solution in warped deformed conifold, which is rather challenging and beyond the scope
 of our current analysis. However, the qualitative agreement between scalings of conformal anomaly indicates we indeed have a
 gravitational description of a QCD like theory.

 \centerline{\bf Acknowledgement}

\noindent I would like to especially thank Keshav Dasgupta for  reading through the draft and making suggestions
 during the course of the work and Miklos Gyulassy for his
valuable feedback. I would also like to thank Martin Kruczenski, Long Chen and Charles Gale for helpful discussions. This work
 is supported by the Office of Nuclear Science of the US
Department
of Energy under grant No. DE-FG02-93ER40764.
 \appendix
\section{Appendix : Three forms and $B_2$} \label{apdx}
Here we explicitly write down the expressions for three forms and their Hodge dual, using the unperturbed metric $\widetilde{g}_{mn}$
  
\bg \label{w2}
\ast_6 \widetilde{\omega}_3^1&=&\frac{3\Gamma_1{\rm sin}\theta_2e^{-B(r)}}{r}dr\wedge\Bigg[\left(1+\frac{6}{9}\left({\rm cot}^2\theta_2-
\frac{a{\rm
cot}\theta_2}{{\rm sin}\theta_2}\right)\right) d\phi_2\wedge d\theta_2
+\frac{6}{9}\bigg(
\frac{{\rm cos}\theta_1{\rm cot}\theta_2}{{\rm sin}\theta_2}\nonumber\\
&-&
\frac{a{\rm
cos}\theta_1}{{\rm sin}^2\theta_2}\bigg) d\phi_1\wedge d\theta_2+\frac{6}{9}\left(\frac{{\rm cot}\theta_2}{{\rm sin}\theta_2}
-\frac{a}{{\rm sin}^2\theta_2}\right) d\psi\wedge d\theta_2\Bigg]
-\frac{3\Gamma_2{\rm sin}\theta_1e^{-B(r)}}{r} dr\wedge
\nonumber\\
&&\Bigg[\left(1+\frac{6}{9}\left({\rm cot}^2\theta_1-
\frac{a{\rm
cot}\theta_1}{{\rm sin}\theta_1}\right)\right)d\phi_1\wedge d\theta_1+\frac{6}{9}\left(
\frac{{\rm cos}\theta_2{\rm cot}\theta_1}{{\rm sin}\theta_1}-
\frac{a{\rm
cos}\theta_2}{{\rm sin}^2\theta_1}\right)\nonumber\\
&& d\phi_2\wedge d\theta_1+\frac{6}{9}\left(\frac{{\rm cot}\theta_1}{{\rm sin}\theta_1}-\frac{a}{{\rm sin}^2\theta_1}\right)
d\psi\wedge d\theta_1\Bigg]
\nd

Using the form (\ref{w2}) and the definition $H_3=dB_2+J_3$, for some $J_3$, we readily find 
\bg\label{B2}
&&B_2=\frac{3g_sM\alpha'}{2}{\rm log}\left(\frac{r}{r_*}\right) \left(g^1\wedge g^2+g^3\wedge g^4\right)+
 12 {\cal K}(r)\kappa_{10}^2g_s{\cal M}N_f\alpha' \mu_7  \Bigg(\Gamma_1{\rm sin}\theta_2\nonumber\\
 &&\Bigg[\left(1+\frac{6}{9}\left({\rm cot}^2\theta_2-
\frac{a{\rm
cot}\theta_2}{{\rm sin}\theta_2}\right)\right)d\phi_2\wedge d\theta_2+\frac{6}{9}\left(\frac{{\rm cot}\theta_2}{{\rm sin}\theta_2}
-\frac{a}{{\rm sin}^2\theta_2}\right) d\psi\wedge d\theta_2\Bigg]\nonumber\\
&& -\Gamma_2{\rm sin}\theta_1\Bigg[\left(1+\frac{6}{9}\left({\rm cot}^2\theta_1-
\frac{a{\rm
cot}\theta_1}{{\rm sin}\theta_1}\right)\right)d\phi_1\wedge d\theta_1+\frac{6}{9}\left(\frac{{\rm cot}\theta_1}{{\rm sin}\theta_1}-\frac{a}{{\rm sin}^2\theta_1}\right)
d\psi\wedge d\theta_1\Bigg]\Bigg)\nonumber\\
&& {\cal K}(r)=\int^r du \frac{ F(u)}{u}\left(1+\frac{\tilde{r}_h^4}{u^4e^{2B}}\right)
\nd


\begin{thebibliography}{}
\bibitem{Witten:1995im} 
  E.~Witten,
  Nucl.\ Phys.\ B {\bf 460}, 335 (1996)
  [hep-th/9510135].
\bibitem{Mal-1}
J.~M.~Maldacena,
  Adv.\ Theor.\ Math.\ Phys.\  {\bf 2}, 231 (1998)
  [Int.\ J.\ Theor.\ Phys.\  {\bf 38}, 1113 (1999)]
  [arXiv:hep-th/9711200].
  \bibitem{Witt-1}
  E.~Witten,
  Adv.\ Theor.\ Math.\ Phys.\  {\bf 2}, 253 (1998)
  [arXiv:hep-th/9802150];
S.~S.~Gubser, I.~R.~Klebanov and A.~M.~Polyakov,
  Phys.\ Lett.\  B {\bf 428}, 105 (1998)
  [arXiv:hep-th/9802109].
   
\bibitem{Gross:1980br} 
  D.~J.~Gross, R.~D.~Pisarski and L.~G.~Yaffe,
  Rev.\ Mod.\ Phys.\  {\bf 53}, 43 (1981).
\bibitem{Pisarski:1981mq} 
  R.~D.~Pisarski,
  Phys.\ Lett.\ B {\bf 110}, 155 (1982).
\bibitem{Appelquist:1981vg} 
  T.~Appelquist and R.~D.~Pisarski,
  Phys.\ Rev.\ D {\bf 23}, 2305 (1981).
\bibitem{Pisarski:1983ms} 
  R.~D.~Pisarski and F.~Wilczek,
  Phys.\ Rev.\ D {\bf 29}, 338 (1984).
  \bibitem{Petreczky:2012rq} 
  P.~Petreczky,
  J.\ Phys.\ G {\bf 39}, 093002 (2012)
  [arXiv:1203.5320 [hep-lat]].

\bibitem{Cheng:2007jq} 
  M.~Cheng, N.~H.~Christ, S.~Datta, J.~van der Heide, C.~Jung, F.~Karsch, O.~Kaczmarek and E.~Laermann {\it et al.},
  Phys.\ Rev.\ D {\bf 77}, 014511 (2008)
  [arXiv:0710.0354 [hep-lat]].

\bibitem{Aoki:2009sc} 
  Y.~Aoki, S.~Borsanyi, S.~Durr, Z.~Fodor, S.~D.~Katz, S.~Krieg and K.~K.~Szabo,
  JHEP {\bf 0906}, 088 (2009)
  [arXiv:0903.4155 [hep-lat]].
\bibitem{Bazavov:2009zn} 
  A.~Bazavov, T.~Bhattacharya, M.~Cheng, N.~H.~Christ, C.~DeTar, S.~Ejiri, S.~Gottlieb and R.~Gupta {\it et al.},
  Phys.\ Rev.\ D {\bf 80}, 014504 (2009)
  [arXiv:0903.4379 [hep-lat]].
\bibitem{Borsanyi:2012ve} 
  S.~.Borsanyi, G.~Endrodi, Z.~Fodor, S.~D.~Katz and K.~K.~Szabo,
  JHEP {\bf 1207}, 056 (2012)
  [arXiv:1204.6184 [hep-lat]].
\bibitem{Panero:2009tv} 
  M.~Panero,
  Phys.\ Rev.\ Lett.\  {\bf 103}, 232001 (2009)
  [arXiv:0907.3719 [hep-lat]].
\bibitem{Panero:2009wr} 
  M.~Panero,
  PoS LAT {\bf 2009}, 172 (2009)
  [arXiv:0912.2448 [hep-lat]].
\bibitem{Mykkanen:2011kz} 
  A.~Mykkanen, M.~Panero and K.~Rummukainen,
  PoS LATTICE {\bf 2011}, 211 (2011)
  [arXiv:1110.3146 [hep-lat]].
\bibitem{Mykkanen:2012ri} 
  A.~Mykkanen, M.~Panero and K.~Rummukainen,
  JHEP {\bf 1205}, 069 (2012)
  [arXiv:1202.2762 [hep-lat]].
\bibitem{Lucini:2012gg} 
  B.~Lucini and M.~Panero,
  arXiv:1210.4997 [hep-th].
\bibitem{Panero:2012qx} 
  M.~Panero,
  arXiv:1210.5510 [hep-lat].
\bibitem{9906194} 
  D.~Z.~Freedman, S.~S.~Gubser, K.~Pilch and N.~P.~Warner,
  JHEP {\bf 0007}, 038 (2000)
  [hep-th/9906194].
\bibitem{9904017} 
  D.~Z.~Freedman, S.~S.~Gubser, K.~Pilch and N.~P.~Warner,
  Adv.\ Theor.\ Math.\ Phys.\  {\bf 3}, 363 (1999)
  [hep-th/9904017].
\bibitem{9909047} 
  L.~Girardello, M.~Petrini, M.~Porrati and A.~Zaffaroni,
  Nucl.\ Phys.\ B {\bf 569}, 451 (2000)
  [hep-th/9909047].
  \bibitem{KS}
  I.~R.~Klebanov and M.~J.~Strassler,
  $\chi_{\rm SB}$-resolution of naked singularities,''
  JHEP {\bf 0008}, 052 (2000)
  [arXiv:hep-th/0007191].
  
\bibitem{Klebanov:2000nc} 
  I.~R.~Klebanov and A.~A.~Tseytlin,
  Nucl.\ Phys.\ B {\bf 578}, 123 (2000)
  [hep-th/0002159].
  \bibitem{Ouyang}
P.~Ouyang,
  Nucl.\ Phys.\  B {\bf 699}, 207 (2004)
  [arXiv:hep-th/0311084].
  
  \bibitem{KT-non-ex}
S.~S.~Gubser, C.~P.~Herzog, I.~R.~Klebanov and A.~A.~Tseytlin,
  JHEP {\bf 0105}, 028 (2001)
  [arXiv:hep-th/0102172].
  A.~Buchel, C.~P.~Herzog, I.~R.~Klebanov, L.~A.~Pando Zayas and A.~A.~Tseytlin,
  JHEP {\bf 0104}, 033 (2001)
  [arXiv:hep-th/0102105].
\bibitem{PandoZayas:2006sa} 
  L.~A.~Pando Zayas and C.~A.~Terrero-Escalante,
  JHEP {\bf 0609}, 051 (2006)
  [hep-th/0605170].
  \bibitem{thorimal}
  O.~Aharony, A.~Buchel and P.~Kerner,
  Phys.\ Rev.\ D {\bf 76}, 086005 (2007)
  [arXiv:0706.1768 [hep-th]];
M.~Mahato, L.~A.~Pando Zayas and C.~A.~Terrero-Escalante,
  JHEP {\bf 0709}, 083 (2007)
  [arXiv:0707.2737 [hep-th]].
\bibitem{Caceres:2011zn} 
  E.~Caceres, C.~Nunez and L.~A.~Pando-Zayas,
  JHEP {\bf 1103}, 054 (2011)
  [arXiv:1101.4123 [hep-th]].
  \bibitem{cotrone}
  F.~Bigazzi, A.~L.~Cotrone, A.~Paredes and A.~V.~Ramallo,
  JHEP {\bf 0903}, 153 (2009)
  [arXiv:0812.3399 [hep-th]];
F.~Bigazzi, A.~L.~Cotrone, J.~Mas, A.~Paredes, A.~V.~Ramallo and J.~Tarrio,
  JHEP {\bf 0911}, 117 (2009)
  [arXiv:0909.2865 [hep-th]];
  arXiv:1110.1744 [hep-th];
A.~L.~Cotrone, A.~Dymarsky and S.~Kuperstein,
  JHEP {\bf 1103}, 005 (2011)
  [arXiv:1010.1017 [hep-th]].
  \bibitem{FEP} 
  M.~Mia, K.~Dasgupta, C.~Gale and S.~Jeon,
  Nucl.\ Phys.\ B {\bf 839}, 187 (2010)
  [arXiv:0902.1540 [hep-th]].
\bibitem{Mia:2009kf} 
  M.~Mia, K.~Dasgupta, C.~Gale and S.~Jeon,
  arXiv:0902.2216 [hep-th].
  \bibitem{Mia:2010tc}
  M.~Mia, K.~Dasgupta, C.~Gale and S.~Jeon,
  Phys.\ Rev.\  D {\bf 82}, 026004 (2010)
  [arXiv:1004.0387 [hep-th]].
\bibitem{Mia:2010zu}       
  M.~Mia, K.~Dasgupta, C.~Gale and S.~Jeon,
  Phys.\ Lett.\ B {\bf 694}, 460 (2011)
  [arXiv:1006.0055 [hep-th]].
\bibitem{Mia:2011iv} 
  M.~Mia, K.~Dasgupta, C.~Gale and S.~Jeon,
  J.\ Phys.\ G {\bf 39}, 054004 (2012)
  [arXiv:1108.0684 [hep-th]].
 
 \bibitem{fangmia}
  M.~Mia, F.~Chen, K.~Dasgupta, P.~Franche and S.~Vaidya,
  Phys.\ Rev.\ D {\bf 86}, 086002 (2012)
  [arXiv:1202.5321 [hep-th]].
\bibitem{Chen:2012me} 
  F.~Chen, L.~Chen, K.~Dasgupta, M.~Mia and O.~Trottier,
  Phys.\ Rev.\ D {\bf 87}, 041901 (2013)
  [arXiv:1209.6061 [hep-th]].

  \bibitem{Mia:2012ue} 
  M.~Mia and F.~Chen,
  JHEP {\bf 1301}, 083 (2013)
  [arXiv:1210.3365 [hep-th]].
  \bibitem{Klebanov:2004ya} 
  I.~R.~Klebanov and J.~M.~Maldacena,
  Int.\ J.\ Mod.\ Phys.\ A {\bf 19}, 5003 (2004)
  [hep-th/0409133].
  \bibitem{kirit} 
U.~Gursoy and E.~Kiritsis,
    JHEP {\bf 0802}, 032 (2008)
[arXiv:0707.1324 [hep-th]];
U.~Gursoy, E.~Kiritsis and F.~Nitti,
  JHEP {\bf 0802}, 019 (2008)
  [arXiv:0707.1349 [hep-th]];
U.~Gursoy, E.~Kiritsis, L.~Mazzanti and F.~Nitti,
  Phys.\ Rev.\ Lett.\  {\bf 101}, 181601 (2008)
  [arXiv:0804.0899 [hep-th]]; 
  JHEP {\bf 0905}, 033 (2009)
  [arXiv:0812.0792 [hep-th]]; U.~Gursoy, E.~Kiritsis, L.~Mazzanti, G.~Michalogiorgakis and F.~Nitti,
  Lect.\ Notes Phys.\  {\bf 828}, 79 (2011)
  [arXiv:1006.5461 [hep-th]].

\bibitem{Tseytlin:1997csa} 
  A.~A.~Tseytlin,
  Nucl.\ Phys.\ B {\bf 501}, 41 (1997)
  [hep-th/9701125].
 
\bibitem{Kuperstein:2008cq} 
  S.~Kuperstein and J.~Sonnenschein,
  JHEP {\bf 0809}, 012 (2008)
  [arXiv:0807.2897 [hep-th]].
  \bibitem{Dymarsky:2010ci} 
  A.~Dymarsky, D.~Melnikov and J.~Sonnenschein,
  JHEP {\bf 1106}, 145 (2011)
  [arXiv:1012.1616 [hep-th]].
  \bibitem{Klebanov:1998hh} 
  I.~R.~Klebanov and E.~Witten,
  Nucl.\ Phys.\ B {\bf 536}, 199 (1998)
  [hep-th/9807080].
  \bibitem{Affleck:1983mk}
  Nucl.\ Phys.\ B {\bf 241}, 493 (1984).

\bibitem{Hawking:1982dh}
  S.~W.~Hawking and D.~N.~Page,
  Commun.\ Math.\ Phys.\  {\bf 87}, 577 (1983).


\bibitem{Witten:1998zw}
  E.~Witten,
  Adv.\ Theor.\ Math.\ Phys.\  {\bf 2}, 505 (1998)
  [arXiv:hep-th/9803131].

\bibitem{wald1}
   R.~M.~Wald,
  Phys.\ Rev.\  D {\bf 48}, 3427 (1993)
  [arXiv:gr-qc/9307038].
  \bibitem{wald2}
   V.~Iyer and R.~M.~Wald,
  Phys.\ Rev.\  D {\bf 50}, 846 (1994)
  [arXiv:gr-qc/9403028].
   \bibitem{wald3}
    T.~Jacobson, G.~Kang and R.~C.~Myers,
  Phys.\ Rev.\  D {\bf 49}, 6587 (1994)
  [arXiv:gr-qc/9312023].
  
\bibitem{wald4}
   R.~Brustein, D.~Gorbonos and M.~Hadad,
  arXiv:0712.3206 [hep-th].
  \bibitem{Meson-spec}
  L.~Chen, K.~Dasgupta, C.~Gale, M.~Mia, M.~Richard and  O.~Trottier, {\it To Appear}. 




  \end{thebibliography}
\end{document}